Equity Default Clawback Swaps to Implement Venture Banking Page 1 of 40

# Equity Default Clawback Swaps to Implement Venture Banking

**Brian P. Hanley**

**Brian.Hanley@ieee.org**

## Abstract

In this theoretical working paper, I propose creation of a venture bank, able to multiply the capital of a venture capital firm by at least 47 times, without requiring access to the Federal Reserve or other central bank apart from settlement. This concept rests on obtaining default swap instruments on loans in order to create the capital required, and expand Tier 1 and 2 base capital. Profitability depends on overall portfolio performance, availability of equity default swaps, cost of default swap, and the multiple of original capital (MOC) adopted by the venture bank.

I propose a new derivative financial instrument, the equity default swap (EDS), to cover loans made as venture investments. An EDS is similar to a credit default swap (CDS) but with some unique features. The features and operation of these new derivative instruments are outlined along with audit requirements. This instrument would be traded on open-outcry exchanges with special features to ensure orderly operation of the market. It is the creation of public markets for EDSs that makes possible the use of public market pricing to indirectly provide a potential market capitalization for the underlying venture-bank investment. Full coverage insulates the venture-bank from losses in most situations, and multiplies profitability quite dramatically in all scenarios. Ten year returns above 20X are attainable.

I further propose a new feature for EDS derivatives which is a clawback lien to close out the equity default swap. Here that clawback is optimized at 77%, and is to be paid back to the underwriter at a future date to prevent perverse incentive to deliberately fail. This new feature creates an Equity Default Clawback Swap (EDCS) which is the proper instrument to use safely.

This proposal also solves an old problem in banking, because it matches the term of the loan with the term of the investment.

I show that the venture-bank investment and the EDCS underwriting business are profitable.

## 1 Preamble

This working paper describes a theoretical system that improves the earnings of venture capital. It assumes two features that must be accepted by regulators. First, that when a loan is made to fund an at-risk real-economy investment (e.g. a standard venture capital convertible note in a real-economy enterprise) that the equity received for that loan principal is not necessarily subject to cancellation per double-entry bookkeeping and is realized as part of leveraged profits because this asset is a real one. There is ample justification for doing this (Hanley 2020). Second, that when an at-risk loan investment from the venture-bank does not return the principal amount, that this is simply forgiven. This has been done by central banks, and I have been told (although I do not have the citation) that this has been done at commercial banks in some instances. Again, there is logical justification for this, as the money that was created by the loan is only owed to the bank that originated it, by the bank that originated it.

Thus, there are two more possible models that could be developed, one which required full cancellation, or some fraction thereof, and another which did not forgive at-risk loans or some fraction thereof.

I have had one discussion with regulators who did not see inherent reason for objecting to these features in the case of truly at-risk investments in real-economy ventures. Most systemic risk occurs when the investments are in finance, and finance of finance to drive exponential bubbles. This has already been shown to be a problem (Hanley, 2012).

## 2 Introduction

Venture capital (VC) seeks to find innovation that will disrupt existing business and by doing so, reap extraordinary returns on investment. Theory relative to venture capital dates back to the 1940's (Schumpeter, 1943). However, the venture capital (debt for equity swap) type of investment appears to date back at least to the ancient Assyrians circa 1800 BC (Gardiner, 2006 pp. 159). However, the ancient Assyrian form was not integrated with modern open-outcry markets. Bankers such as J.P. Morgan were major investors in Edison and Tesla, performing the venture capital function in the late 19th and early 20th centuries, however, today, banking is not generally directly involved in venture capital transactions.

Banking, contrary to what most people believe (including some regulators classically trained), does not lend out money the bank has on deposit. Instead, banks create deposits by writing loans (McLeay, 2014). This is crucial to grasp in order to understand this venture banking proposal.

### 1.1 Venture-capital today

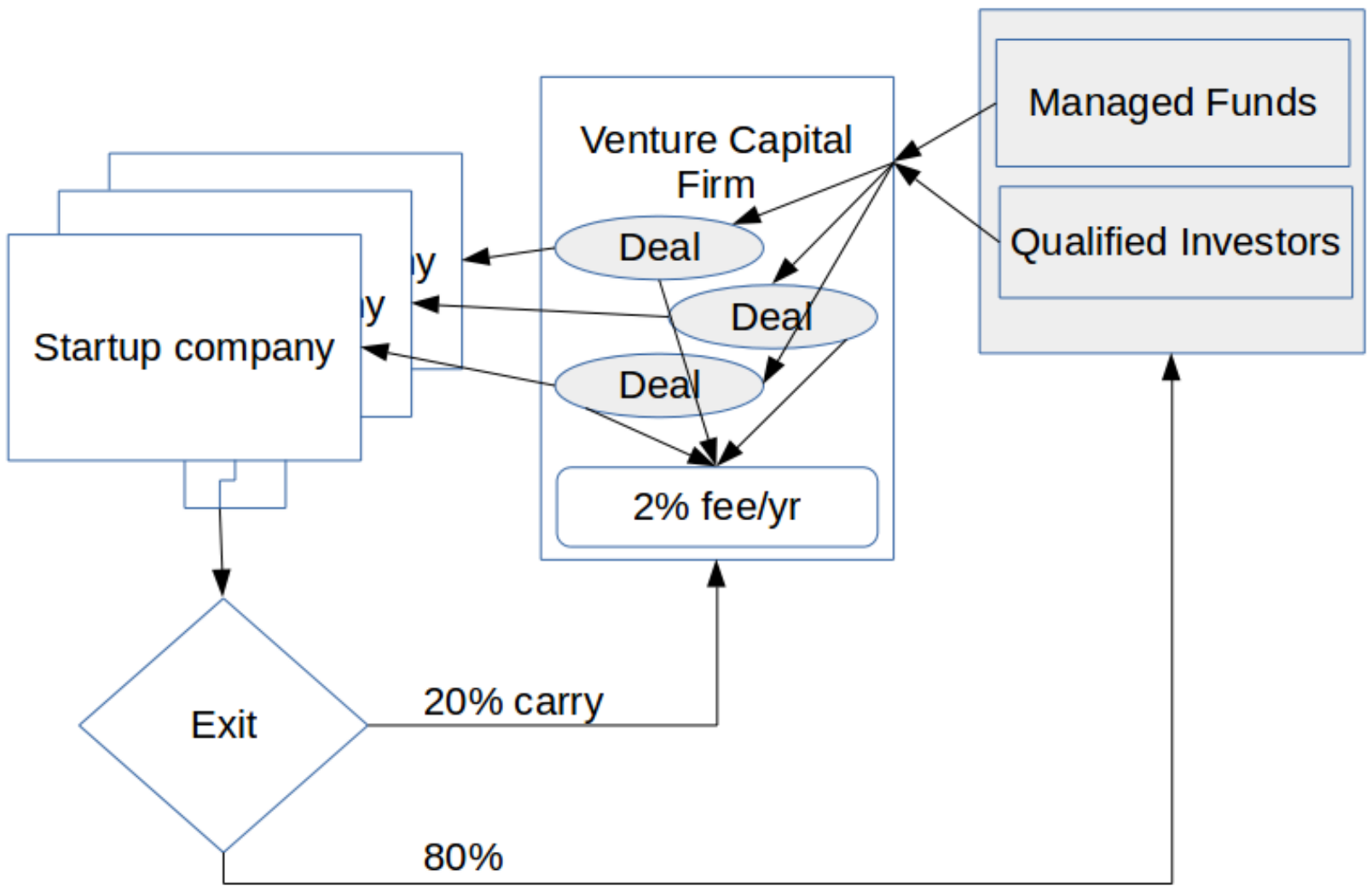

*Figure 1: Venture capital diagram*

There are a range of venture capital firms which invest in seed, startup, early, or late stage of development of a company prior to exit. Some VC firms perform all stages, some specialize in a specific stage and work with other firms for later or earlier phases. Venture capital today participates in money creation at exit, when the shares of an investment are priced on the open market. Until then, either through initial public offering, an acquisition or other private sale, the valuation is not realized.

Venture capital firms today solicit limited partners (LPs), typically pension funds, endowment funds, and other large capital management funds. Many such institutional LPs have fixed fractions they commit to venture capital to balance their portfolios. Some qualified investors also sign up as LPs. With such investor commitments, the VC firm looks for ventures to invest in. Generally, the VC firm will request that money be transferred from an LP only when a deal is ready, as their clock starts when LP funds are transferred. While many VC firms attempt to go from investment to exit in 5 years, a 10 year cycle is more realistic.

With rare exceptions, venture capital today is dominated by VC firms where partners in the firm contribute 1% or less of funds under management. A few funds are operated from the money of partners alone. The compensation structure of a typical VC fund pays fees of 2% of fund assets per year to the firm. When investments are closed out, 20% of the profit goes to the VC fund, which is referred to as "carry". (Mulcahy, 2012)

### 1.2 Historical investment banking, a prelude to venture banking

In the Schumpeterian disruption ethos of Silicon Valley's venture capitalists, banking is arguably the greatest disruptive innovation our world has ever seen. By the innovation of banking, capital was multiplied many-fold. Over time, this radical innovation that created money on demand through debt, has developed into modern money. Banking replaced hard money reserves with central banking, and hence makes banking demand driven only limited by qualified borrowers. With the development of modern monetary theory an accounting identity is recognized between debt balances and asset balances (Visser, 1991). For every credit balance on account, there is an equal and opposite debt balance, and vice versa. Some would argue with this, but there is really no question that all credit balances originate in a debt balance somewhere. This is what a loan is – it is a note declaring that money now exists, and since in an accounting system, money is just accounting entries, this works. This is the innovation that the Medici family discovered and codified as a formality. It is doubtful that the Medici's discovered it for the first time, as there are indications that Rome may have had some form of what we call banking (Temin, 2004). It is also possible that certain ancient societies had something along the lines of banking, as debt/credit money appears to predate hard money tokens (Graeber, 2014; Quiggin, 1949).

In the late 19th and early 20th centuries, investment banks were used as a vehicle for wealthy investors that functioned similarly to what a venture capital firm does today. Such banks pooled the capital of a group of partners, and then made their investments as loans to real enterprises, using their capital as the reserves of the bank. This gave them the leverage that banking creates. However, over time, what investment banks did expanded and evolved into the current day investment bankers, whose largest single activity has been finance of finance (Korten, 2009, p.21). Serious abuses of banking leverage that fueled the stock market crash of 1929 led to regulation that essentially ended banks as venture-capitalists. This proposal defines a new form of banking that directly finances real-economy new ventures. But in concept it harks back to a century ago.

### 1.3 Kraken banking money creation

***A.*** *A bank issues a loan to a borrower.*

***B.*** *The bank purchases a derivative to act as insurance against default on the loan.*

***C.*** *The bank then books this insured valuation of the loan into its capital account.*

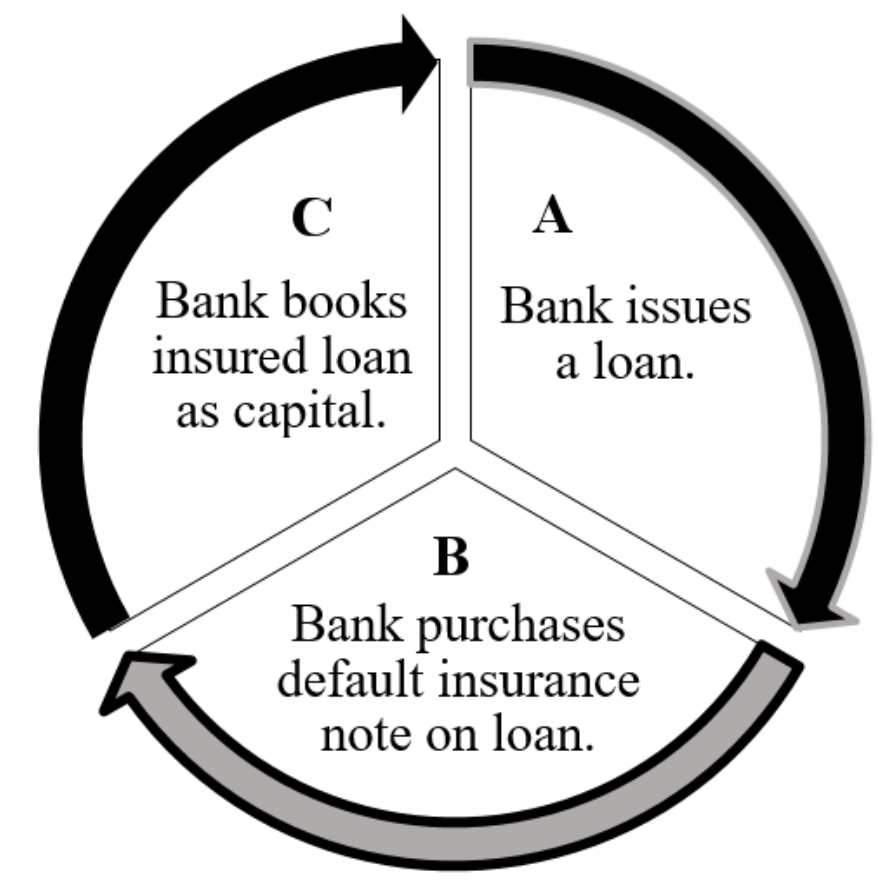

*Figure 2: Money creation cycle of Kraken banking*

During the years of run-up to the global financial crisis of 2008, an emergent phenomenon appeared that had both positive and negative aspects. It was a heretofore unknown algorithm for creating money in the banking system (Hanley, 2012). I named this emergent phenomenon the Kraken, and it's enabling method the Default Insurance Note (DIN). The term DIN is a class of instrument that encompasses CDS contracts, the Equity Default Swap (EDS) contract that I conceived, and the final version which is the Equity Default Clawback Swap (EDCS) contract.

This new money multiplier produces logarithmic curves that do not converge. What the logarithmic graph in figure 3 shows is that capital available is only capped by real-world limitations on capacity when investment-loan insurance underwriting is available.

The Kraken curves in figure 3 are described by equation (1) below (Hanley, 2012).

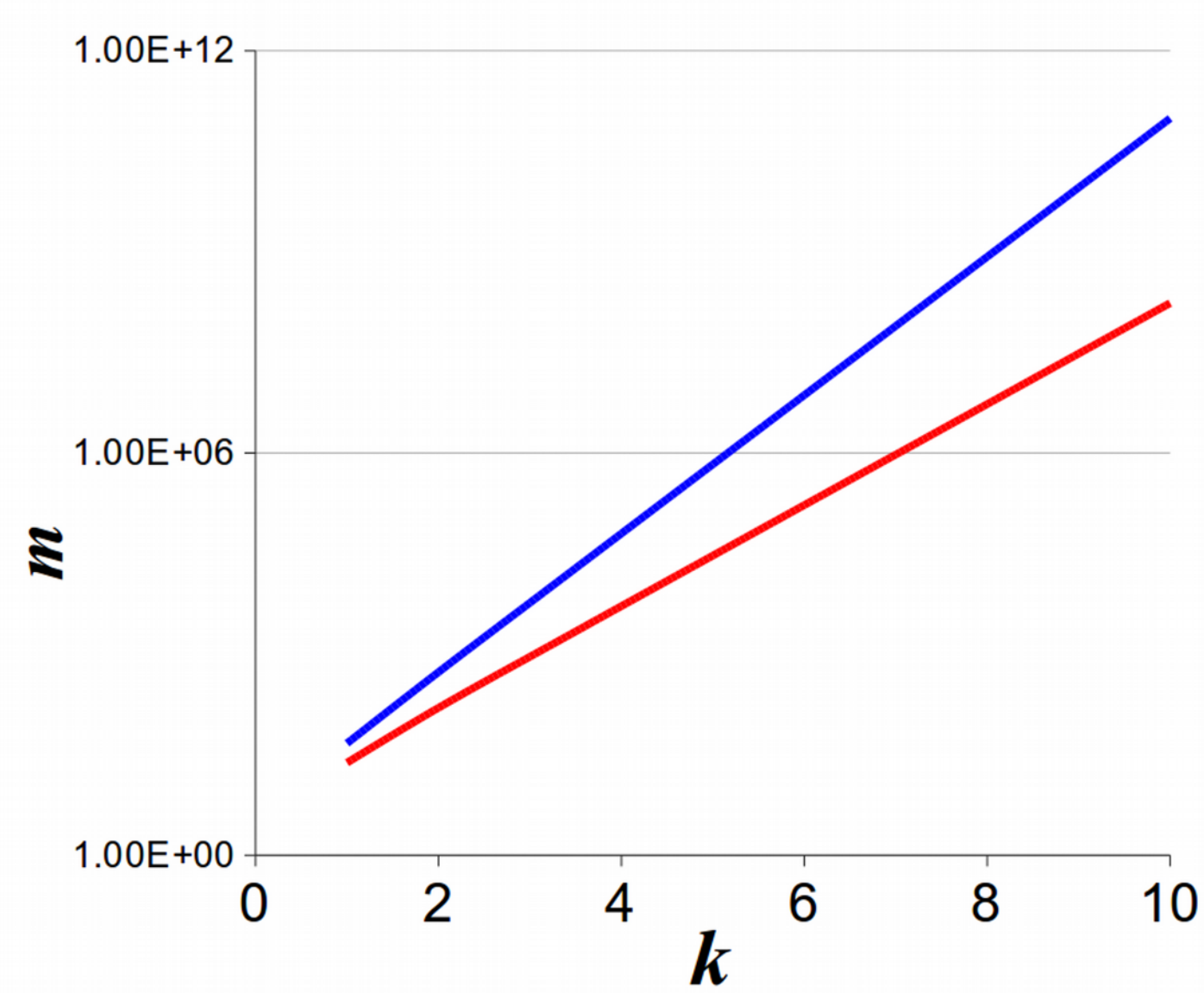

*Figure 3: Kraken curves, log scale. Shows values of* ***m****, for* ***k*** *= 1 to 10. (Red, lower line)* ***R****=5%, (Blue, upper line)* ***R****=2.5%. (****R****=Reserve ratio) Reproduced from fig. 3 (Hanley, 2012).*

$$m=\sum_{i_1=0}^{n}\left((1-R)^{i_1}+\left(\left((1-R)^{i_1}(O-I)T\cdot\sum_{i_2=0}^{n}\left((1-R)^{i_2}+\left(\left((1-R)^{i_2}(O-I)T\cdot\ldots\sum_{i_k=0}^{n}(1-R)^{i_k}+(1-R)^{i_k}(O-I)T\right)\right)\right)\right)\right)\right) \quad (1)$$

*Where:* ***R*** *= deposit reserve fraction,*

$i_1, i_2 \ldots i_k$ *= iteration number on loans/deposits,* ***k*** *being the series end term*

***n*** *= iteration limit,* ***I*** *= insurance price as fraction*

***O*** *= 1 + origination fee fraction of loan (generally charged as "points")*

***T*** *= tranche fraction insured.*

### 1.4 Classical banking money creation review

Contrast the log-scale behavior of figure 3 with the classical banking multiplier's asymptotic behavior in figure 4, which has **1/*R*** as its limit, where ***R*** is the reserve fraction. Thus, in classical thinking, a reserve ratio of 5% yields a multiplier asymptote of 20 times reserves. In the real world, banks make loans based on their Tier 1 or Tier 2 capital (Bank, 2016) which forms the reserves they have. Part of Tier 1 capital belongs to the bank's principal owners, and is at risk. So, a newly formed bank can make loans based on its capital reserves, and doesn't need any deposits.

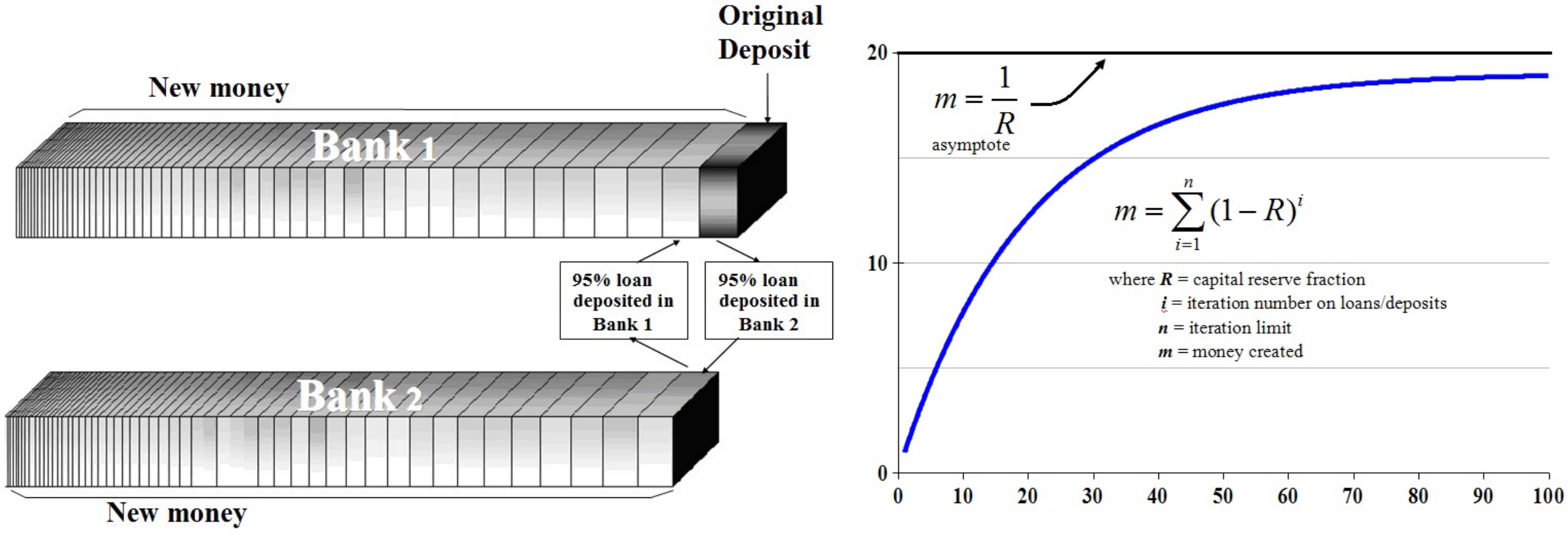

*Figure 4: Classical banking multiplier that assumes some original reserve money. Typically, this would be gold or silver, although in reality, banks violated reserve relationships long ago.*

In the modern world of banking, things are more flexible. In theory, unlimited capital is available from the central bank. A bank can make a loan and has 2 weeks to settle. Reserves can be borrowed from, or assets sold to, the central bank after loans are made. However, access to those loans and easing is dependent on the central bank (for instance, the Federal Reserve) oversight. In practice banking is limited by regulations that tend to direct bankers into hard assets rather than value creation. This is quite odd the more one studies economics, because manufacturing and services (that people need or want) is where "real economy" value-creation occurs (Schumpeter, 1943). Hard assets like real estate inflate in price without creating new value once they are built. So we see banks as both enablers of rent extraction, and as rentiers themselves. Rentiers charge for access to an asset without creating any value. Venture-capital falls into a category considered high-risk, even though risk is actually quite acceptable on broad portfolios.

Almost half a century before the creation of central banks such as the Federal Reserve, Thomas Tooke reported that gold reserves no longer functioned as they were supposed to in the Bank of England, and the bank had little practical interest in gold or silver. (Tooke, 1844) This provided a foundation to the later formalization of the abandonment of precious metals, and replacement with the central bank system that could provide as much reserve as necessary.

### 1.5 Basel accords and insured asset capital

The exponential mechanism shown in figure 3 is possible because of regulation that allows insured assets as capital (Division, 2016), which is consistent with Basel accords. This mechanism can, if properly applied, make increased capital available, up to the limit of the real-economy to utilize it[1]. This represents a secondary method of money creation that bypasses central banks as arbiters of reserves.

### 1.6 BACPA and derivative instruments resolution

A significant feature of using derivative instruments to insure loans is that the owner of the derivative can have instant resolution in case of default on the note. First appearing in the United States in 1982, safe-harbor bankruptcy code provisions put the claims of derivative holders first, granting them the right to terminate and complete transactions immediately upon bankruptcy of a counterparty (Gilbane 2010). Thus, derivative holders have the right to immediate foreclosure on underlying assets. These safe-harbor protections cover forward, commodity, and security contracts; and also cover repurchase and swap agreements. All other claims in bankruptcy are given an automatic stay and must wend their way through the courts. With the passage of public law 109-8, or BACPA, in 2005 (Grassley 2005), these provisions were clarified and strengthened to ensure newer instruments would be covered.

### 1.7 Misapplication of Kraken mechanism

What came to light in the aftermath of the 2008 GFC was an emergent problem with AIG–Major Banks (e.g. CitiGroup, Bank of America, et al.). First, that this logarithmic growth mechanism existed; second, that it had been misapplied – aimed at a sector (real estate) that did not create new utility value along with increasing price valuation (Hanley, 2012); and third, that a method by which the Federal Reserve implicitly monitored activity did not work with this new mechanism because the banks didn't need to go through the Federal Reserve to replenish reserves in order to make loans. I believe that development of the Kraken banking mechanism was accidental, an unplanned emergent phenomenon that came about between the major banks and underwriters offering CDS contracts – AIG primarily.

---

1 The "real economy" is the productive economy of goods and services that have utility value to people. Pushing money into the real economy beyond it's ability to create utility value creates a bubble to one extent or another.

## 3 Venture banking

This proposal assumes no access to a central bank's reserve mechanism. As discussed above, the asymptotic classical banking multiplier (1/*R* where *R* = reserve fraction) no longer operates in the way assumed in classical banking because of the invention of reserve banking around the turn of the 20th century. (e.g. Federal Reserve Bank, Bank of England, European Central Bank, etc.). However, this venture-banking proposal is designed to function without access to a central bank except for settlement purposes. This is intentional, in part because this is a new concept and banking regulators are conservative, so it will take time before venture-banking is viewed in the same way other banking is. Consequently, it was required that I discuss multiplier dependent banking above.

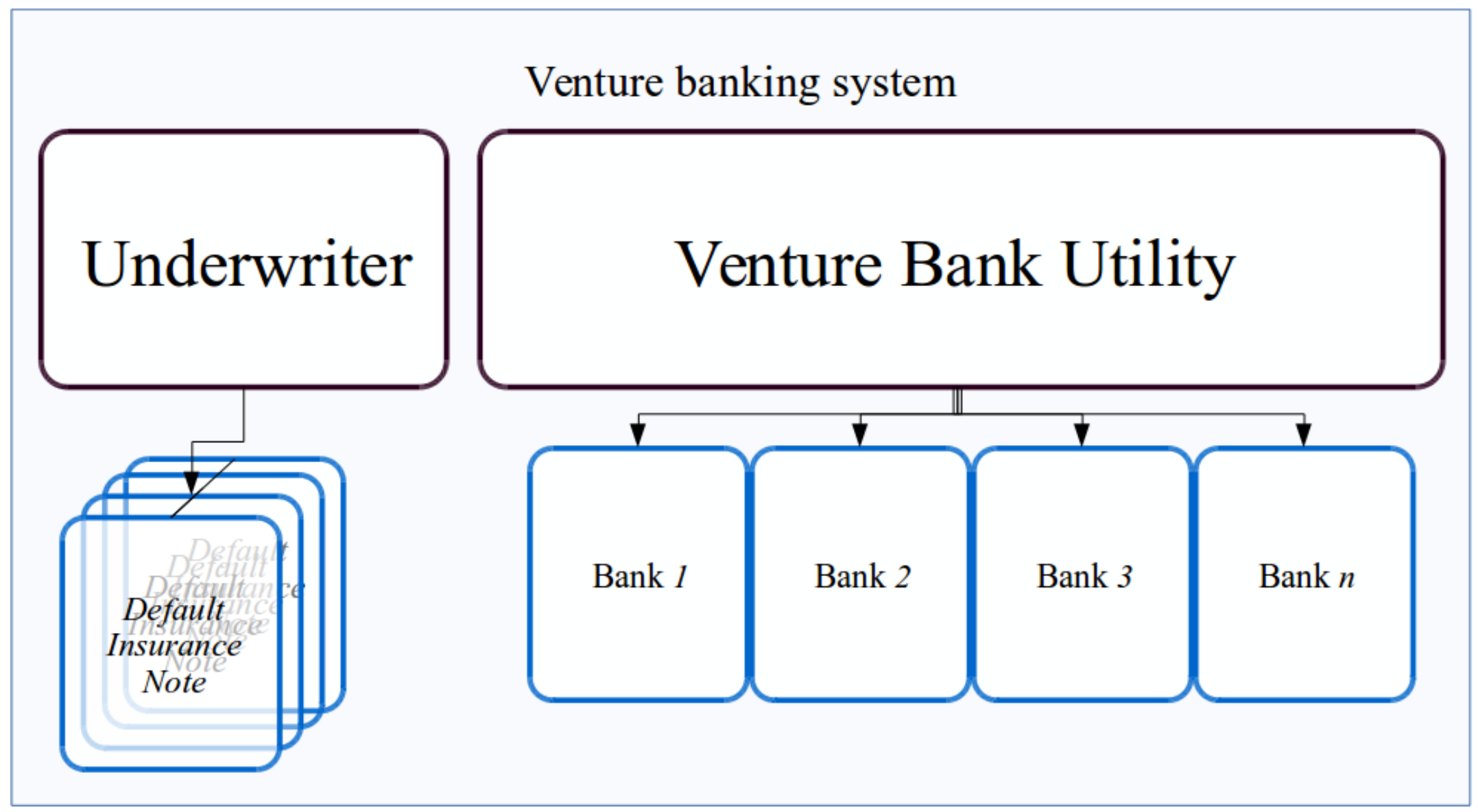

*Figure 5: Proposed venture capital banking system. Venture bank utility (VBU) is paired with an underwriter issuing Equity Default Swaps (EDCSs).The clients of the VBU are venture capital firms.*

I propose that a new venture bank utility (VBU) be created as a utility service instead of each venture-bank organizing itself separately. The VBU's services would be paid for by fees assessed on the venture bank sub-entities. The concept is similar to that of Mastercard, which was created as a system owned by its member banks, solely to process credit card transactions[2]. I do not, in this proposal, account for such fees.

This venture bank utility (VBU) system, diagrammed in figure 5, is intended to simplify the process of opening a bank for use by venture capital firms. The way this would work is that an

2 It may be wise to diversify this banking mechanism in multiple countries in an effort to prevent complete dependence on the regulators of one country.

investor or group of investors would deposit their money into the VBU, and the utility would handle the overhead of operating as a bank.

An underwriting entity parallel to the VBU would write the derivative contracts to insure the loans made by the investor's banks. This could be a department of the VBU, or it could be a separate entity depending on the regulatory environment where the VBU operated. Equity Default Clawback Swaps (EDCSs) effectively insure as much of the investment loans as needed by the venture banks.

In my conception, a VBU would operate as a not-for-profit corporation owned in partnership by the venture-capital firms that were members. This would require joining fees and the member venture banks would participate in a similar manner to the way current commercial banks have run the Mastercard clearing house.

The creation of a VBU system has some challenges, but fundamentally the VBU is a matter of execution. The major questions arise for the EDS derivatives, because it is necessary to define how these instruments will work, and it is the EDS that is the key to this novel banking variant.

### 3.1 Winnowing of Venture Banks

Unlike stock picking on exchanges, the past performance of venture capital funds appears to be correlated with future results (Mulcahy, 2012). Consequently, it is worth considering that a well run VBU might perform periodic evaluations that I would expect to begin around 5-10 years. The object of these evaluations would be to decide whether a specific venture bank should be wound down, and removed from future participation in the VBU due to poor performance. Figure 7 shows returns for the large Kauffman venture capital portfolio. While some degree of losses can be tolerated if each of these funds were operated as a venture bank within a VBU, ending future relationship with proven poor performers on the lowest decile of figure 7's graph may improve future earnings for the VBU as a whole, and provide places for new VC firms to participate that do not have a track record.

However, winnowing is not necessary, and there are valid arguments against it. Primary among those arguments is that as we will see, venture banking can tolerate significant average losses (~50% ) on investments and still stay in the black, making similar returns to venture capital today (Hanley, 2018). What this means is that ventures are possible that would otherwise not be. In the long run, this could benefit society.

### 3.2 Regulatory limits on Venture Banking

In the United States, the regulations most pertinent are mostly contained in section 2020.1 of the Commercial Bank Examination Manual of the Federal Reserve (Division, 2016). These regulations change from time to time based on legislation and any agreements created by the Basel Committee. The Basel Committee is hosted by the Bank for International Settlements (BIS) (Basel, 2016). Nations participating in the international banking system have regulations conforming to BIS similar to US regulations.

For banks that are members of the Federal Deposit Insurance Corporation (FDIC) system, requirements are more stringent. Disclosure that the bank is making equity investments is required, and current regulations recognize that this activity is more common. The securities involved here would be equity investments, and, as such, bookable as Tier 1 capital at fair value to the extent they are realized. **A significant element of my proposal is how to create a credible realization mechanism through public market valuation of venture bank investments prior to exit. I think the EDCS could potentially perform that function, because it represents a kind of futures contract.**

The most common category of venture bank securities would be Type III, and these should not exceed more than 10% of capital to any single obligor. For this proposal, obligors greater than 10% should not generally be the case. A venture bank should also be able to create portfolios that would be Type V securities which can go to 25% of capital for each such security.

> *"The supervisory guidance in SR-00-9 on private equity investments and merchant banking activities is concerned with a banking organization's proper risk-focused management of its private equity investment activities so that these investments do not adversely affect the safety and soundness of the affiliated insured depository institutions."* (Division, 2016)

The focus of compliance for this venture bank entity will be on developing broad portfolios and showing regulators, through models and records, that the bank will not create a systemic risk hazard. My view is that a well run venture banking system will significantly lower overall risk, not raise it. A VBU system should also drive professionalization of venture-capital institutional operations[3].

---

3 What exactly would be good practices for venture capital is a topic all its own. It is an art, not a science, and includes factors such as connections, overall intelligence and intellectual openness and creativity that are hard to quantify. But there are factors such as technology viability, systems to cure problems, market size and reach that do have solid metrics.

### 3.3 Limits to the money creation multiplier of a EDCS-VBU system

The Basel accords specify some critical elements for capital requirements in banking that are relevant here. Tier 2 capital is in superior position to Tier 1 in liquidation; 15% of the Tier 1 capital and 100% of Tier 2 capital can be insured assets. Tier 2 capital cannot exceed Tier 1 capital, giving simple equations 2 and 3 that show a 47X limit on outstanding money creation.

$$\frac{C}{0.85} = Tier1\,limit \rightarrow \frac{2C}{0.85} = Capital\,reserves\,limit \quad (2)$$

$$\frac{1}{R}\left(\frac{2C}{0.85}\right) = Maximum\,loan\,limit \quad (3)$$

*Where:* ***C*** *= Initial capital,*

***R*** *= Reserve percentage.*

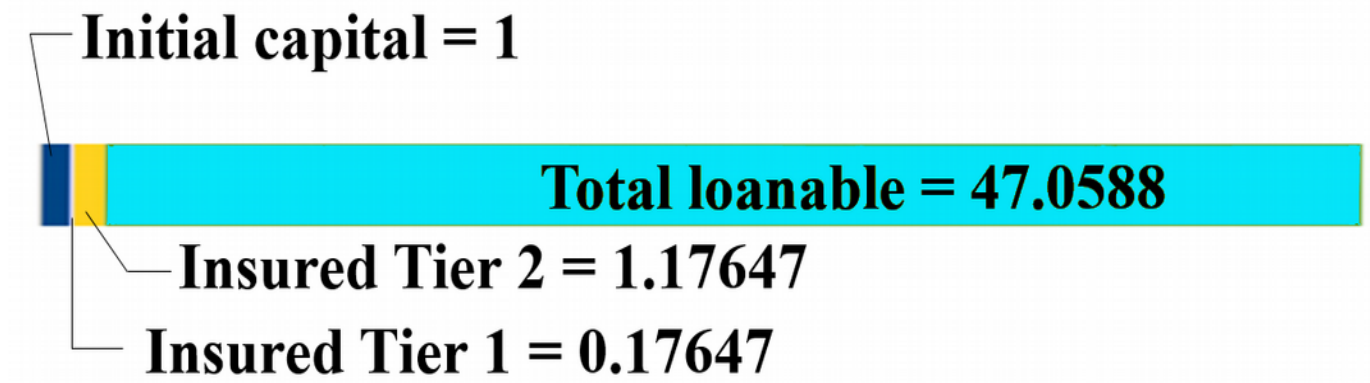

*Figure 6: Venture bank capital (Tier 1 & 2) and loan limits graph for R = 5%*

The 47X limit is the outcome of regulation. Those regulations were created to ensure that investors in the bank have sufficient assets at stake to operate the bank properly. Venture-banking is different from ordinary banking, however, because the primary stake for venture capitalists is in the investments themselves. Since this would be equally true, if not more so, in venture-banking than in current venture capital firms, (the rewards are higher) I think that in the long-term venture banks could have such regulatory limits minimized safely. This should be safe as long as the system is operated properly.

This simple limit of 47X is easily exceeded inside a 10 year period within the model I have created which assumes investment failures will occur at the 5 year mark, and exits will occur at 10 years. Assuming that 75% of investments don't return capital (Ghosh/Gage, 2012) and are written off in 5 years, then for the second 5 year period, another 35X is available for a total of 82X. Assuming that investments exit at 10 years, then at 10 years another 25X is available, plus 75% of the 35X, for a total after year 11 of 47 + 35 + 25 + 26 = 133X. Again, this presumes that venture banks would not have access to central bank reserves.

### 3.4 Handling of bank deposits created by venture-bank investment-loans

The bank's demand deposits would be composed of both deposits made into it by its supported companies when an investment loan is written, and by revenue collected by the companies invested in. Assuming that early companies would have burned through half of their allocated funds at any particular time, and that later stage companies would be making money, a perhaps reasonable assumption is that there would be half the invesment loans deposited in the venture bank present at any time. On average, ¼ of the deposits on hand would be reasonably expected to be available for short-term loans, and such interest that the bank paid to itself to finance its insurance premiums would either be used for operations, or else booked into Tier 1 capital to enlarge the total loan limit.

### 3.5 Optimum system—100% EDCS coverage

This system is analogous to the mechanism between AIG and the major banks where real estate loans were insured. Those loans became insured assets, hence, for the term of the insurance contract, the value of the insurance policy is assigned to the capital account, and is available for a new loan. This provides an overwhelming surplus of Tier 1 and 2 eligible capital. In this proposed system, if you do the math, only 2.88% of the EDCSs needs to be present to provide enough Tier 1 and 2 capital. The rest is accounted as if it were long-term savings accounts, and available to loan. However, in this proposal, 100% of the loan amounts are insured to satisfy regulators that the funds are completely safe, as an alternative to "finding the money" to cover outstanding loans which can prove quite expensive, and exposes the rest of the banking system to serious risk.

### 3.6 Outline of an EDCS

The primary difference between an EDCS and a CDS contract is that the EDCS includes transfer of equity to the underwriter when the underlying investment is sold or goes public. So for the seller, there are two items of value in the transaction, the payments received from the buyer and a claim for some amount of equity in the investment. There are two equity claims on the investment by the EDCS:

1. Triggers can give 100% of equity to an EDCS holder. Compared to a CDS, a trigger corresponds to the default on a loan but is more complex, as will be discussed.
2. IPO or sale (e.g. M&A) of the investment gives a negotiated percentage to the venture bank, which in the models in this paper is 50% of exit equity. (e.g. at exit, 50% of the exit equity is paid to the venture bank.)

There are also other provisions of EDCSs that are discussed in more detail below, however, the primary benefits to the underwriter/seller are the premium stream and the investment equity. Both premiums and equity fractions can vary.

In a simple example, if the EDCS equity is 50%, and the EDCS has covered 100% of an investment into a company, and when the company had it's IPO, this investment owned 20% of the IPO stock, then the EDCS equity would be: 20% investor equity x 100% EDCS coverage x 50% of EDCS = 10% of the total equity in the IPO. The venture bank would receive the other 50%. There is a cost for using this mechanism, but that balances against the increase in funds available to use, and the higher return that leverage based on money creation creates because of the difference between a true loss and a write-down.

### 3.7 Requirements for participation in EDCS banking

To participate in a VBU, it would be necessary for venture banks to transparently show their investments for use by underwriters. Optimally analysis based on these records would be public, and this information should be compiled by an exchange-related entity for use by underwriters and for other analysis. Using these records from a pool of investment cash flows and exits, the premiums required can be calculated. I see these records becoming a key part of the system for setting rates, for both sides of the EDCS transaction.

A primary regulatory concern would be ensuring that the EDCS underwriters would stay solvent under virtually all circumstances. A single venture bank in a VBU can fail and not crash the system. But failure of an underwriter could have repercussions that include bringing down all of the investments it has insured. I could see a period where, similarly to AIG's CDS business unit, an insurance company could be motivated to price their instruments below true solvency as a way of taking market share and generating paper profits. It would be wise to prevent this, either by industry agreement, or by legislation. It is my intent that EDCS premiums be flat across the industry.

Many parties, institutional and private, should qualify for direct purchase of entire EDCS contracts, and shares in EDCSs could also be sold to individuals much the way that stocks are. I would suggest that regulators should wish to see part of the EDCS volume sold to the general public be shares in a portfolio. In theory this should work, as long as underwriters are prevented from dumping bad investments as bundled securities while keeping the good ones for themselves. Dumping bad investments could be prevented through regulation requiring that either all of the underwriting from an insurer (or group of insurers) for a predetermined period going forward, or a randomly chosen set of *n* unsold EDCS contracts chosen by a third party for an insurer, would be packaged into a security. This should prevent underwriters from using insider knowledge to make the public market unfair, which is a serious temptation.

## 4 Venture bank modeling

### 4.1 Net return data source – Kauffman's set of 99 venture capital firms

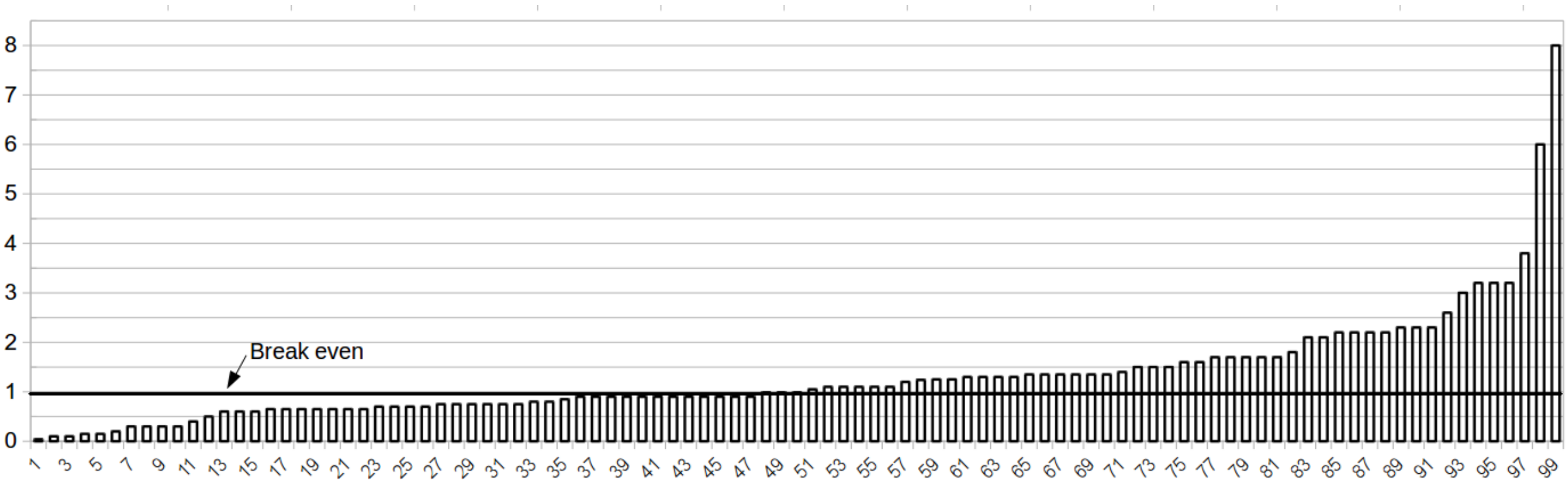

*Figure 7: Kauffman venture capital fund dataset. (Mulcahy, 2012)*

The risk model is based on the published data from the 99 venture capital firms Kauffman invested in. Returns were compiled over 20 years for the 10 year performance shown in Figure 7. Each bar is the return on investment for one of the 99 venture capital firms. All returns shown are net of each VC firm's 2% fee and 20% carry compensation structure. My model uses a compressed version of this with the same overall characteristics obtained by use of a fitted curve equation. By adding and subtracting from this dataset, varying rates of return are modeled, all with the same overall spread as Kauffman's dataset. The model allows adjustments that lower and raise returns.

### 4.2 Portfolio models with no LIBOR borrowing, 100% EDCS coverage and 5% EDCS premiums

*4.2.1 High return portfolio 1.50 – Based on Oregon Public Employee Ret. Fund*

| EDCS rate | **Classical Portfolio return** | EDCS Underwriter investment (using year 5 as payout year) | EDCS Underwriter 10 year premium earnings | EDCS 10 year Net profit | EDCS Underwriter 10 year return. (1.00 = break-even) | EDCS Underwriter yearly return | EDCS Equity fraction |
|---|---|---|---|---|---|---|---|
| 5% | 1.50 | -4.82 | 54.07 | 49.25 | 11.22 | 27.35% | 50% |

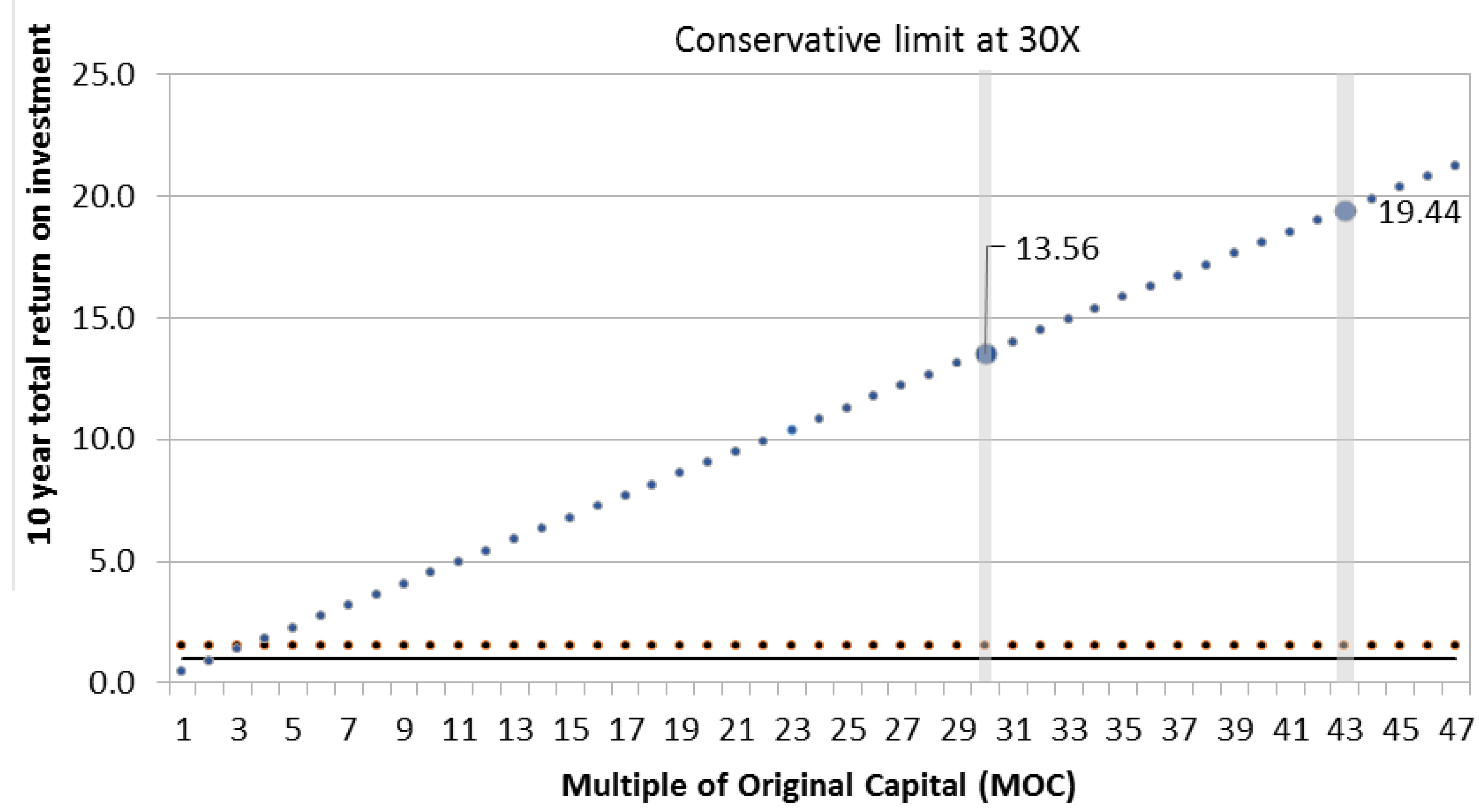

*Figure 8: Booking EDCS value to capital account where a classical return would be 1.50.*

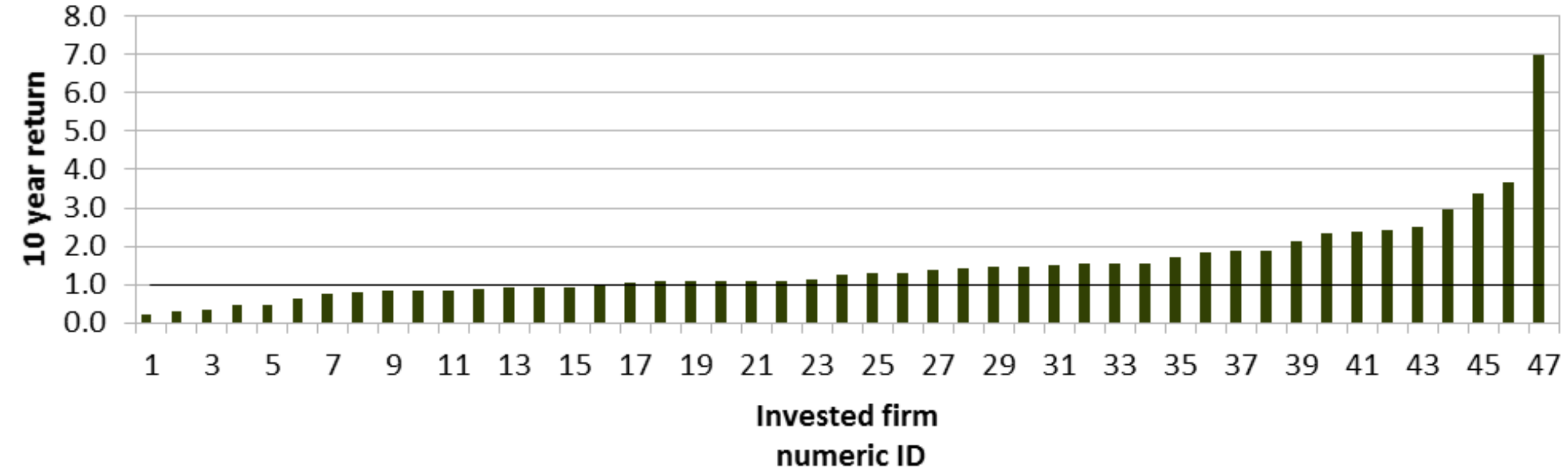

*Figure 9: Venture funds distribution of 10 year returns for a 1.5X total portfolio. 1 = breakeven*

*4.2.2 Intermediate return portfolio 1.31 – Based on Kauffman funds portfolio*

| EDCS rate | **Classical Portfolio return** | EDCS Underwriter investment (using year 5 as payout year) | EDCS Underwriter 10 year premium earnings | EDCS 10 year Net profit | EDCS Underwriter 10 year return. (1.00 = break-even) | EDCS Underwriter yearly return | EDCS Equity fraction |
|---|---|---|---|---|---|---|---|
| 5% | 1.31 | -8.43 | 50.14 | 41.71 | 5.95 | 19.52 | 50.00% |

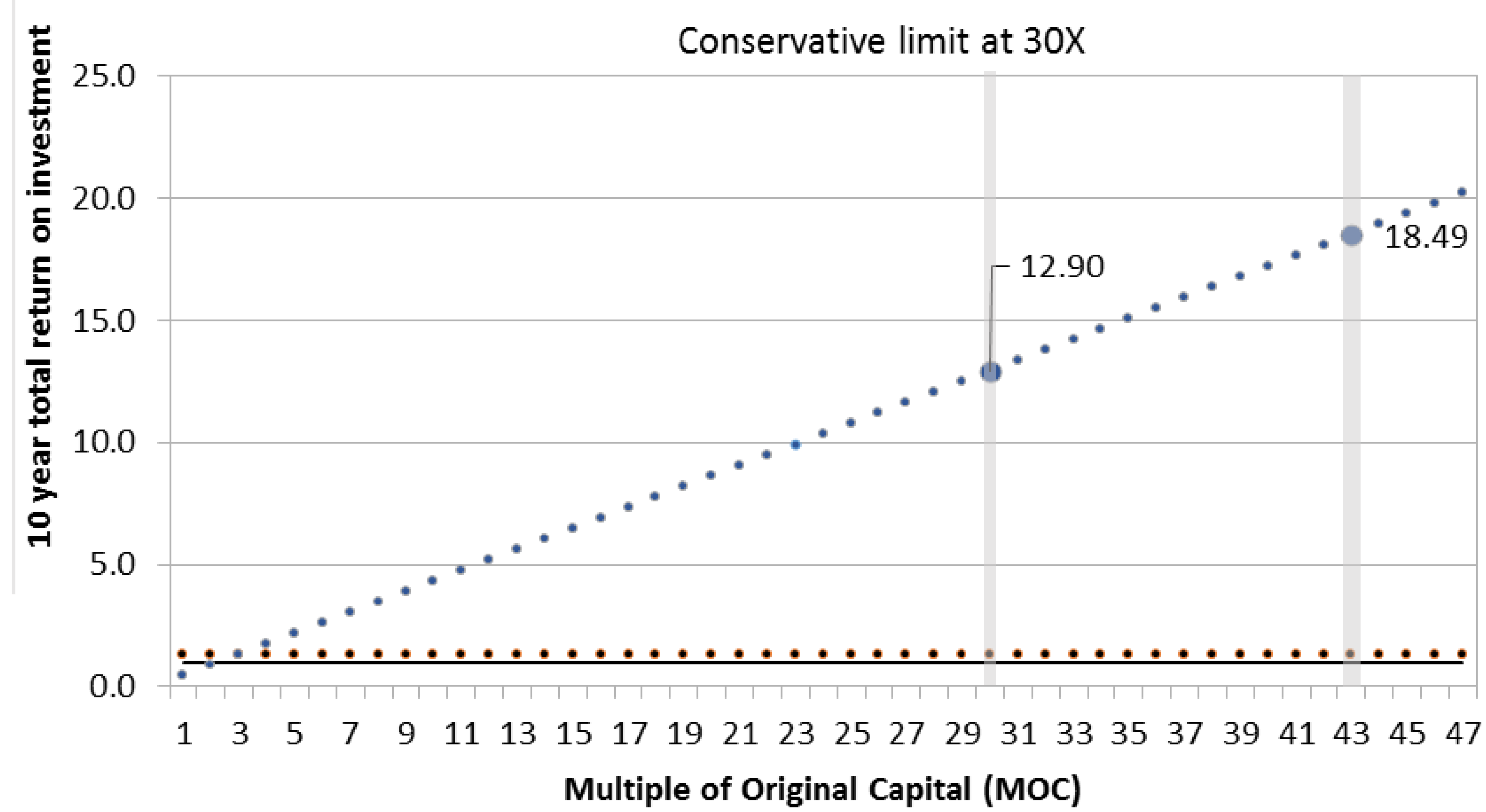

*Figure 10: Booking EDCS value to capital account where a classical return would be 1.31. Note that a little less money is made here than at either 1.10 or 1.50 classical portfolio's return. This is because of EDCS payoffs.*

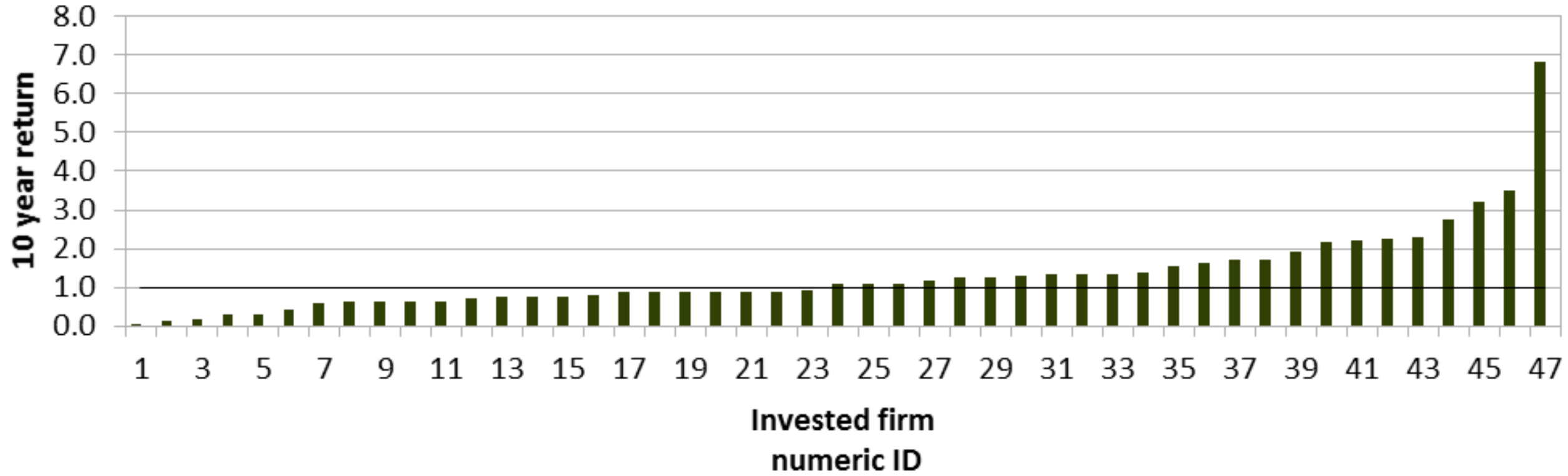

*Figure 11: Venture funds distribution of 10 year returns for a 1.31X total portfolio. 1 = breakeven*

*4.2.3 Low performance portfolio 1.10 – Based on Prequin median net multiple returns bottom tertile, all private equity, 2000-2015. 100% EDCS coverage, 5% EDCS premium*

| EDCS rate | **Classical Portfolio return** | EDCS Underwriter investment (using year 5 as payout year) | EDCS Underwriter 10 year premium earnings | EDCS 10 year Net profit | EDCS Underwriter 10 year return. (1.00 = break-even) | EDCS Underwriter yearly return | EDCS Equity fraction |
|---|---|---|---|---|---|---|---|
| 5% | 1.1 | -13.84 | 44.23 | 30.39 | 3.2 | 12.32% | 50.00% |

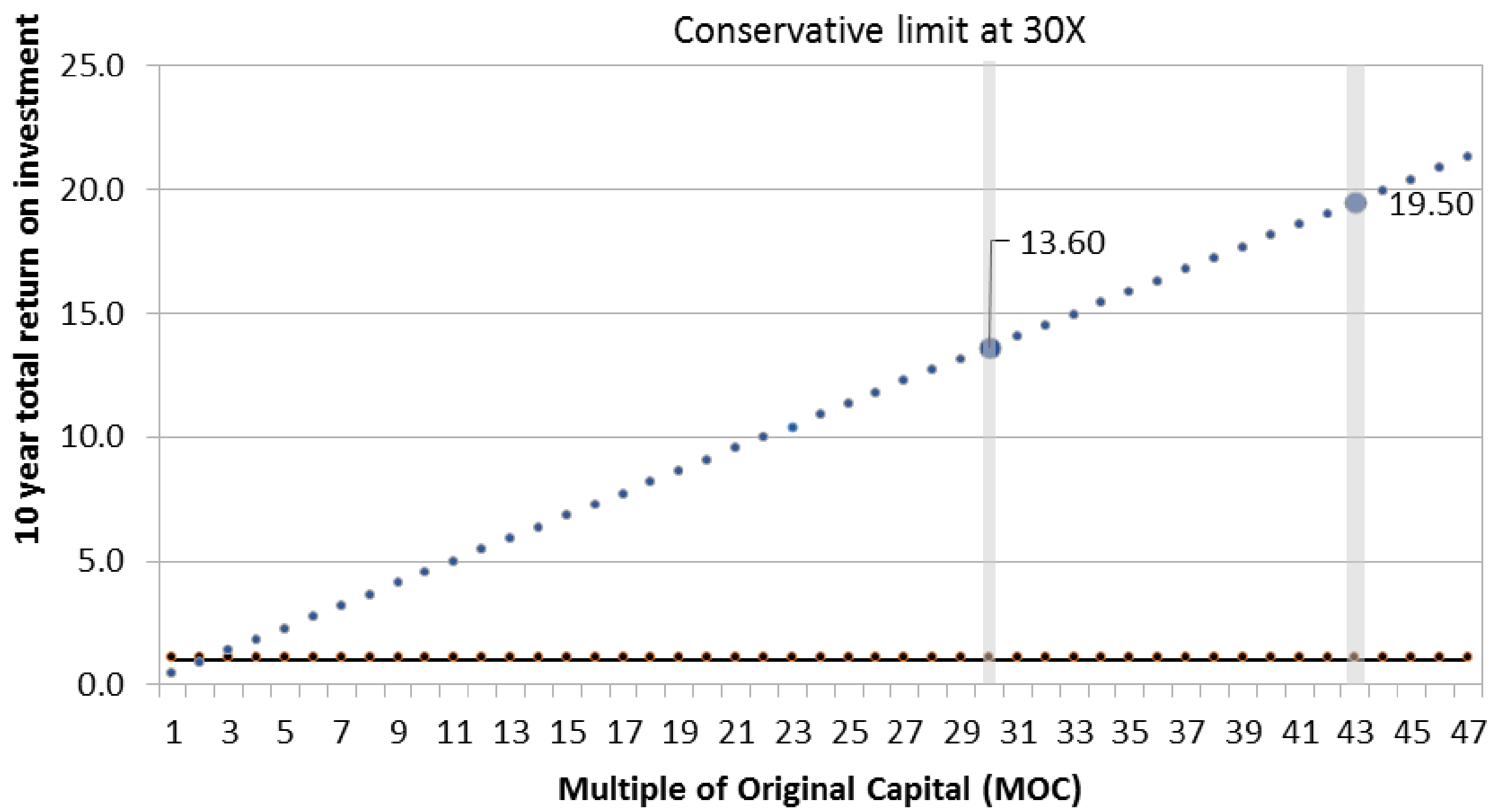

*Figure 12: Booking EDCS value to capital account where a classical return would be 1.1. Eventually, as returns drop, the EDCS underwriter loses their shirt, so EDCS premiums may benefit from some flexibility, perhaps including post-hoc adjustments under certain scenarios.*

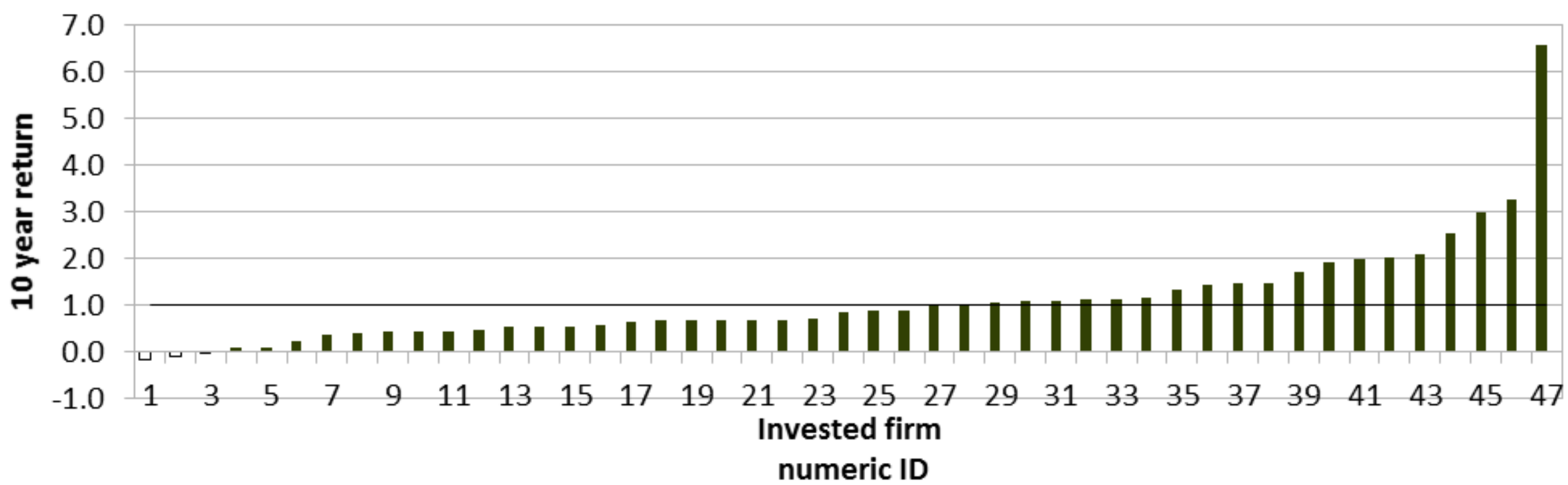

*Figure 13: Venture funds distribution of 10 year returns for a 1.10X total portfolio. 1 = breakeven*

### *4.3 Equity Default Swap graphs without clawback*

#### *4.3.1 100% EDS without clawback: Relationship of total 10 year classical return to venture-bank returns*

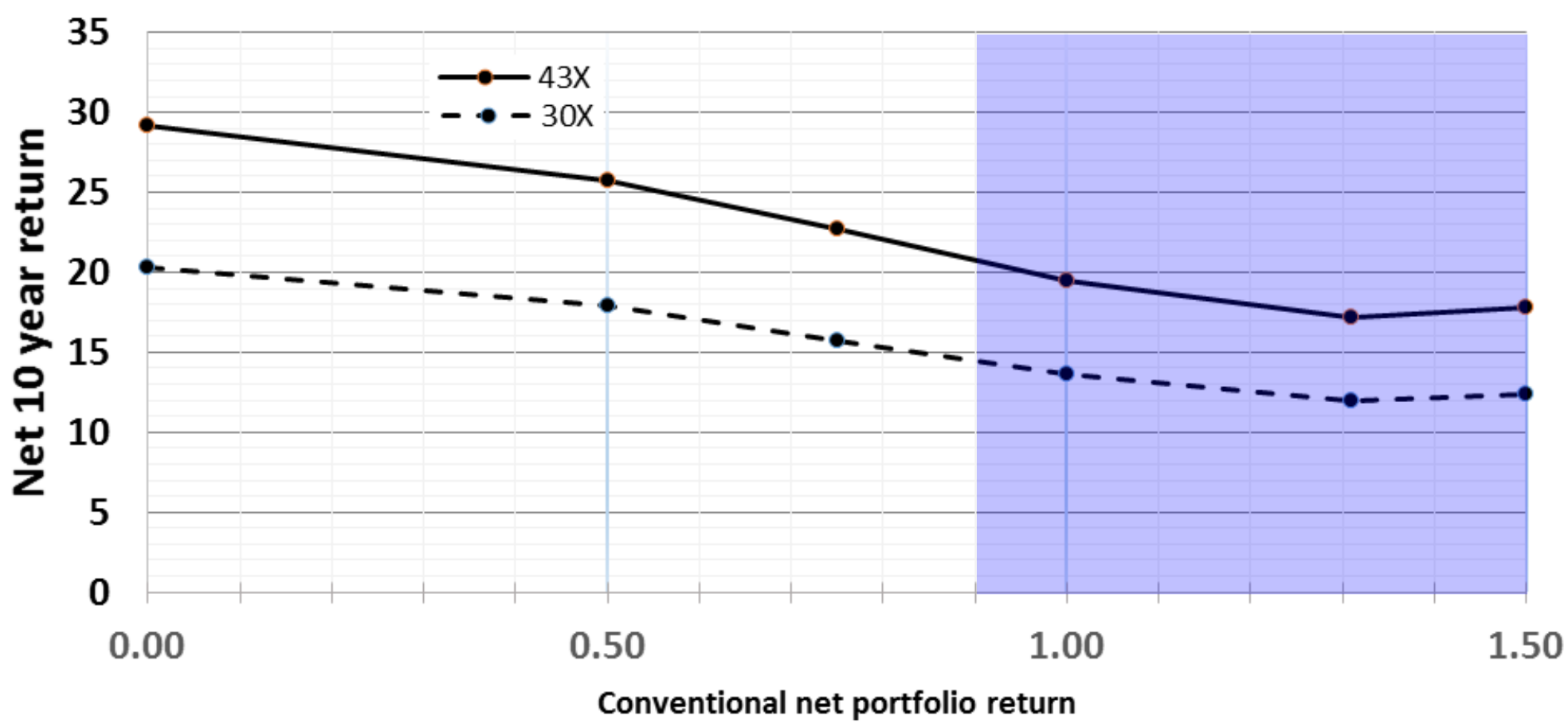

*Figure 14: At 5% EDS premium and 50% equity for 100% EDS coverage without clawback.* ***Break-even = 1.0.*** *Shows net 10 year total return varying what the return of a conventional portfolio would have been. Shading indicates normal range of returns for large VC portfolios. Graphs for 30X and 43X MOC. Break-even at 1.0. A conventional total return cutoff of 1.50 was selected because this was the highest actual large portfolio found. Above 2.0 however, returns become quite high.*

What we see here is that returns are protected due to insurance protecting capital, and returning the book value of each venture investment loan that fails. The shape of this graph shows a perverse incentive for the venture bankers. Since the EDSs are derivatives and immediately enforceable, they cannot be litigated or qualified in the way of normal insurance.

EDS return is calculated by dividing total return by the amount of payoff money paid out, adjusted for time and cost of money. This shows that at the low end, the simple formula of 5% per year premiums and full payouts, with the same level of equity sharing do lead to losses. But in the more normal range returns are very good. It is necessary to improve the EDS formula to ensure EDS returns are strong even at the low end. This is done by having a clawback lien to close out the instrument that makes up losses, while still satisfying regulators that the loan assets are fully insured. The clawback lien creates a steadily rising curve for the venture-banks as discussed separately in "The perverse incentive for insurance instruments that are derivatives: solving the jackpot problem with a clawback lien for Equity Default Swaps" (Hanley 2017). This final derivative instrument is the Equity Default Clawback Swap (EDCS).

*4.3.2 Relationship of total 10 year classical return to 10 year EDCS return*

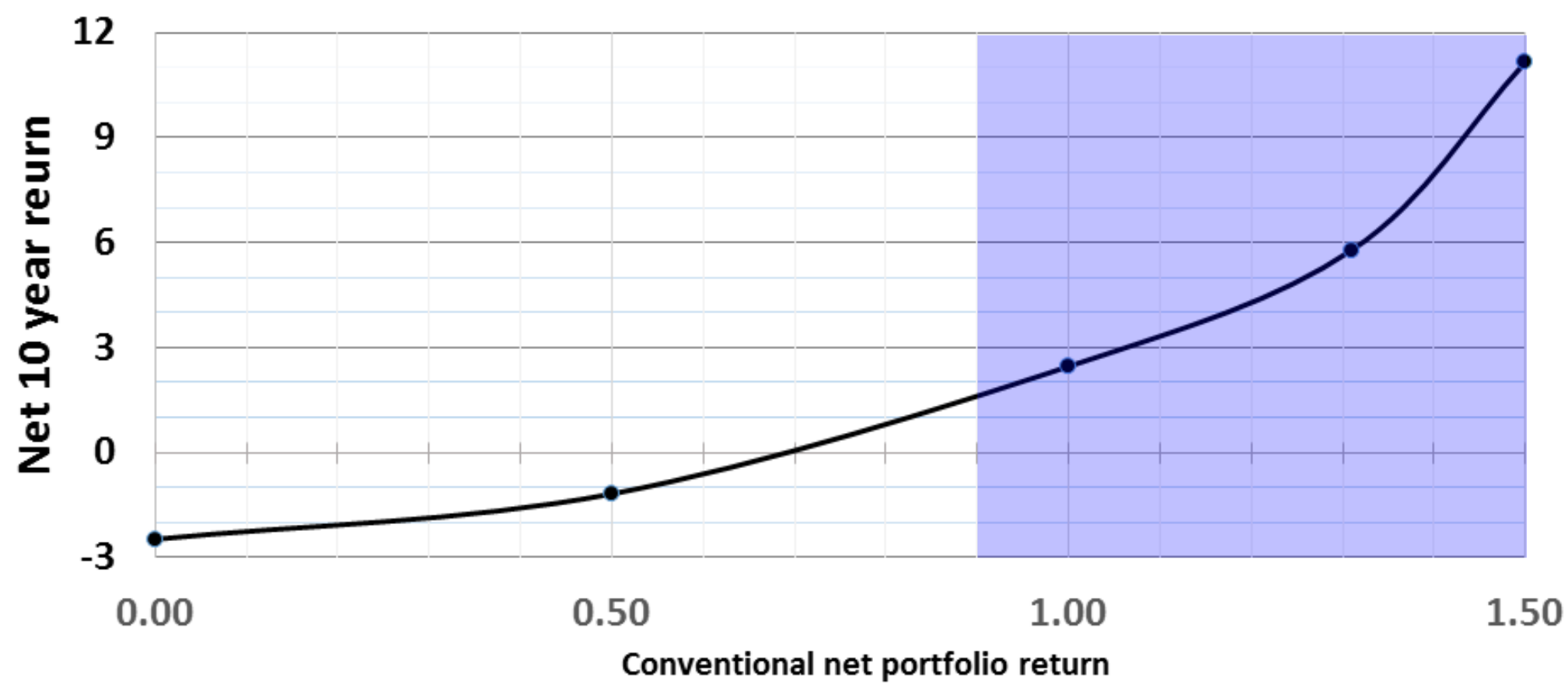

*Figure 15: EDS underwriter's return without clawback.* ***Breakeven = 0*** *Since the average VC fund returns approximately 1.0 over 10 years, this graph indicates a good business, provided that the firms the underwriters take over are liquidated in a manner matching to the Mulcahy data.*

# 5 EDCS contracts – Market, structure, and operations.

## 5.1 Equity Default Clawback Swap derivatives market

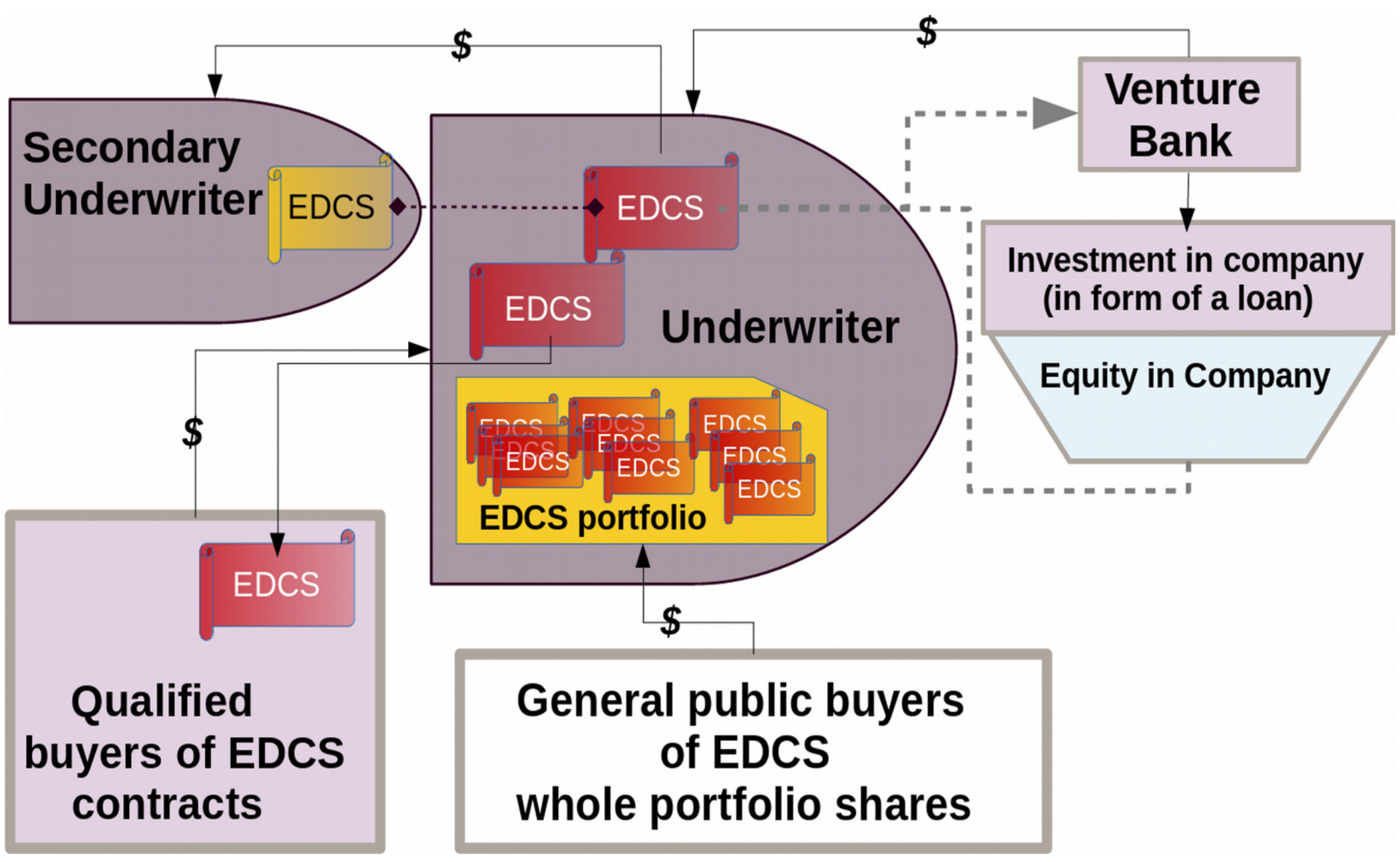

*Figure 16: EDCS derivatives market diagram. The interactions with buyers that apply to primary underwriters can also apply to secondary underwriters, provided that audit requirements are complied with.*

The primary counterparties to an EDCS contract are venture bankers who buy EDCSs in order to cover some fraction of their loans (generally 100%). EDCSs are potential assets suitable for transfer in the event a venture capital group decides to dissolve. When triggered due to closeout of the equity at a loss, the EDCS is an instrument that transfers the EDCS's equity share of the company in return for a cash payment by the insurer.

A robust market could include secondary parties who could be interested in buying the underwriter's side of the EDCS for investment purposes. EDCS instruments are also a way for the general public to participate in long-term venture capital investments. This market for secondary parties may not be strictly required. There is no open outcry market for CDS contracts – it is all off-exchange. However, this is against the opinion of regulators (Weistroffer, 2009), and it is my opinion that an open outcry exchange market for the underwriter side of EDCS contracts is critical.

## 5.2 Care in defining EDCS contracts so they are practical

A key aspect of any derivative instrument such as the EDCS is that any metrics need to be black and white, easily seen as true or false within defined time windows. Law dictates the speed with which these instruments are exercised and prevents litigation from taking place. Therefore it is necessary to avoid overly complex clauses that leave room for subjective interpretation. Otherwise, the courts might get involved and rule that such an instrument was no longer properly classed as an ordinary derivative. That, in turn could cause the collapse of the whole system.

I present what I believe to be a set of contract provisions enforceable as derivatives. Consider this strong guidance.

### 5.3 Open-outcry public market

An open market provides for transparency and fairness. An open market makes it possible for the general public to participate along with institutional investors and provide more liquidity. And an open-outcry market creates a mechanism by which the underlying stock value of non-public companies could be valued. This mechanism could provide an acceptable anchor for venture-bank loan activity by means of a market-based valuation, provided that underwriters are prevented from withholding what they know to be the best investments and selling off the worst performers. An open outcry market is favored by regulators (Weistroffer ,2009).

### 5.4 EDCS concepts

The two-party scenario assumes that there is an issuer (the underwriter), and that the buyer (the venture bank) would only have one interest, getting paid in the case of a trigger. The underwriter's side expects to collect premiums and then equity at exit, or else pay off the cash value of the note to the buyer when a trigger occurs and take over 100% of assets underwritten.

Triggers are defined in detail below. Potential triggers are: Non-payment of premium, bankruptcy of investment, turning down a bona-fide offer to buy, and gain or loss exit from the investment.

The original buyer side of the EDCS coverage loosely corresponds to a put in options trading. It gives the holder the right to “put” the underlying assets to the EDCS holder and get paid at an agreed upon price when the investment is exited with a loss. The original underwriter's side of the EDCS coverage similarly corresponds to a call in options trading. It gives the holder the right to obtain equity or equivalent cash when an investment is exited with a gain.

From here on, I will refer to these as the put and call sides of the EDCS. However, unlike standard put and call contracts, these would be the two sides of the same contract. The put side is the original buyer, the investing bank. The call side is the underwriter, who collects premiums and equity.

Other differences from normal put contracts is that the put side of an EDCS could not be sold or transferred separately from the underlying loan, and normally not sold at all by the venture bank except in extraordinary circumstances, like winding down an entire VBU or creating a pseudo-exit if the EDCS went beyond 10 years.

### 5.5 EDCS premiums and term

A typical EDCS would pay a premium annually, with a minimum term of 5 years, and a nominal term of 10 years with an option to extend annually by the purchaser, if needed[4]. It is necessary that EDCSs be extendable, or that alternative forms of exit be available because evidence from venture capital shows that in some VC firms venture holdings exist as long as 17 years later. An EDCS premium level would be pre-negotiated, not subject to unexpected changes by the issuer nor subject to yearly renegotiation should a buyer's market develop[5]. For the underwriter, it is necessary to prevent a convexity trap developing in which the venture bank renegotiates on ventures that go well, and holds on ventures that go badly. This may be an area for regulation to intervene to protect the industry from itself.

To cancel the EDCS and remove the encumbrance on the equity of the investment, a payment would need to be made, negotiated with the seller of the contract. But normally, the purchaser of the EDCS must pay their agreed upon premiums, equity, or cash in lieu of equity, to the EDCS issuer. The nominal premium used in my modeling is 5% per annum, and the equity returned to the underwriter is 50% of the EDCS coverage. Lower premiums can be used with increased exit equity, or if the market shows over time that this is practical. However, the lower the premium, the higher the carrying cost of payoffs, which cuts into exit profitability for the venture bank. Premiums higher than 5% can become problematic, and 5% or less is adequate for good financial performance .

I set a minimum of 5 years of premiums in my modeling. This prevents buyers from purchasing EDCS contracts and closing out their investments at a loss in the first year. The proper functioning of the clawback at the rough 75%-77% fraction of realized loss for the underwriter can still be gamed if the time window of insurance is short enough.

---

4 In practice, an EDCS could be priced on a yearly, quarterly, monthly or some other basis. In my modeling, for simplicity, I used 12 month pricing, basing this on LEAPS contracts. The requirement that matters is that, once issued, the policy cannot be canceled, and the pricing is predictable within some fairly narrow limits.

5 An exception to pre-negotiation could occur if a state regulator emerges. This is what would happen if the parties could enter negotiations each year (or possibly 2 or 5 years) on the price of a EDCS. Because the EDCS would have to be purchased, and underwriters could have short term incentive to underprice either the back end equity or the premiums, an outside regulator could be necessary to set rates.

### 5.6 Investments held for more than 10 years.

As seen in figure 17, an issue is EDCS contracts that extend out beyond the 10 year window of my current modeling. I could conceive of a EDCS which would raise the exit equity for each year beyond 10 years in order to compensate for EDCS carrying cost, possibly at a lower rate, however this has not yet been modeled. Alternatively, it could be possible for a pseudo-exit in a very long-term investment to occur, in which the venture bank, or the firm invested in, pays off the underwriter at a level both consider acceptable, perhaps with an arbitrator. Another alternative as this system matures could be selling these assets to a central bank. I could conceive of a EDCS contract that put an upper limit of 10 or 15 years on the term, and required a valuation followed by payment in such cases. There are, after all, completely viable firms that have excellent reasons for staying private, and there are mechanisms such as bonds for raising money to execute an internal buyout. The overall goal should be that the EDCS business be protected from losses, and venture-banks be given as much latitude to operate as reasonably possible.

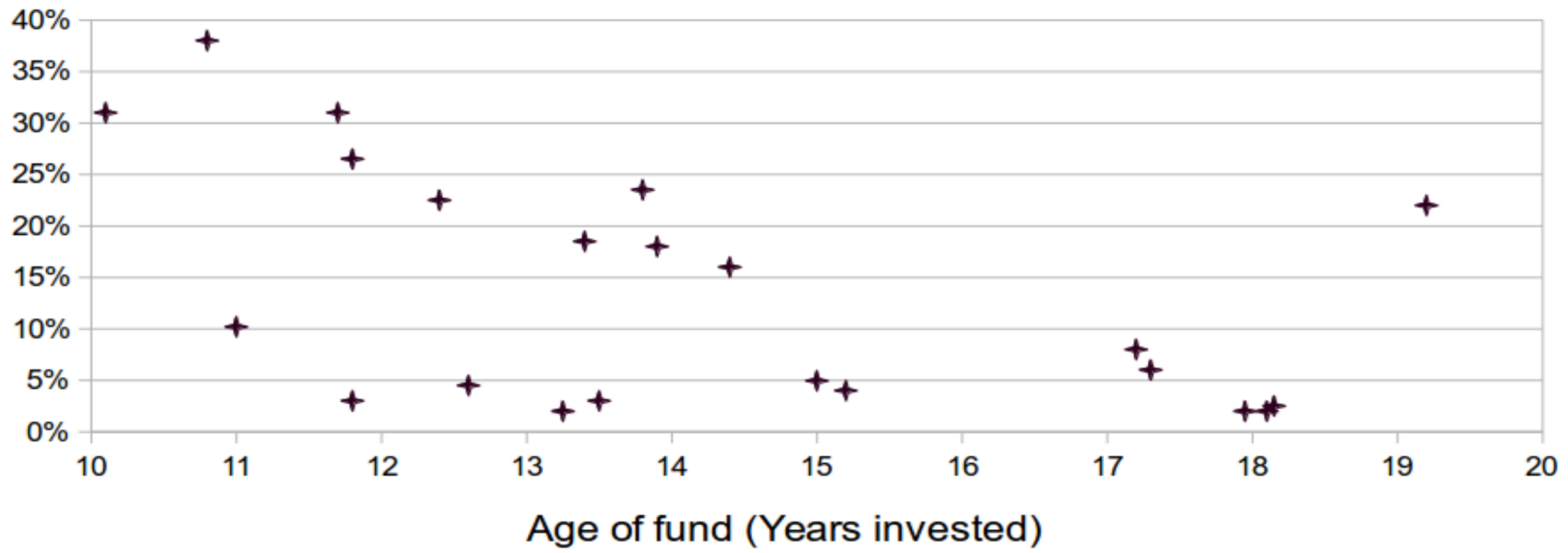

*Figure 17: Kauffman—Percentage of non-liquidated assets in a subset of venture funds. 22 out of 99 of these funds had assets past theire 10 year target. (Mulcahy, 2012)*

#### *5.6.1 Packaging investment portfolios*

Investment portfolios should be set up that package a group of investments. For instance a set of venture bank loan-investments over a 1 year period would be grouped and accounted for as a set as they mature. When an EDCS payout to a venture bank occurs for an investment in that set, that payout would normally be held by the venture bank for the balance of the agreed upon time period required to close out the insured portfolio. After closeout of the portfolio, the clawback lien amount would be paid back to the underwriter. I would expect the normal close out period to be 10 years. However, as seen above, some fraction of investments do not exit on schedule. Some exit early, and some exit late. Should investments exit late, then the alternative of folding those into a set of rolling portfolios may be helpful.

In this alternative, a venture bank's assets would be managed as a single, rolling portfolio of sub-portfolios. Bona-fide ventures that are unable to exit on schedule could be moved into another portfolio, and a substitute of a similar venture in terms of monetary valuation that has a closer exit date could replace it. This could potentially also be a way to turn back or reset the clock on extended term ventures. If this were done, then EDCS payouts would generally be released collectively as fractions as the rolling sub-portfolios closed out. In general, I am in favor of as much flexibility as possible for both sides when that flexibility improves the overall stability of the system and improves scheduled liquidity. Time squeezes are an enemy of stability and orderly markets.

### 5.7 Types of EDCSs and exchange registration

All EDCS contracts will be registered on an exchange that records the deal they cover, the amount and terms. This does not necessarily need to be all public information, but it should be available to underwriters, venture banks, and analysts. Regular reports should be publicly available, and the data should be freely available to any analyst that is bona-fide and willing to abide by whatever privacy provisions are decided upon. EDCS contracts will be a new type of futures derivative that allows public participation in both the future new ventures and in the soundness of the venture bank's management of their portfolio. Thus, it will be necessary to provide for analysts that can properly inform the public while also protecting confidentiality of ventures to the extent necessary.

Primary EDCS – Issued by an underwriter and sold to the original investor, typically a venture capital banker to insure a loan. This EDCS registers the underlying equity investment, the banker, and the loan.

Secondary EDCS – A second underwriter may issue a secondary EDCS referencing the primary EDCS held, and sell it to the issuer of the primary EDCS. When the primary EDCS is triggered, this propagates back to the referencing EDCS. Consequently, it is required that a secondary EDCS reference its primary, and that the primary reference it's secondary. Having these references recorded on the exchange makes it possible to evaluate the true outstanding value of EDCSs, and determine which party is the ultimate beneficiary. I address this matter because CDS contracts have operated this way in an effort by banks (which acted as underwriters) to spread risk. However, this has resulted in notional value of CDS contracts that are multiples of the loans they insure and there is no way to track what is real and what is not.

A regulator could decide that secondary EDCS contracts should not exist for the sake of simplicity. However, as above, I generally favor flexibility if it could improve stability. That is the primary question that should be answered. If there are secondary EDCS contracts, will they improve stability and scheduled market liquidity, while lowering system risk?

### 5.8 Fraud and EDCS in an adversarial relationship

There is an inherent adversarial relationship between venture bankers seeking EDCSs and underwriter-issuers of EDCSs. An obvious fraud strategy is to package portfolios of bad investments and sell them off to unsuspecting suckers, while retaining the EDCSs. This way the fraud perpetrator gets paid off if the suckers lose – after recovering their capital in the sale. This strategy was pursued by the major banks against AIG in the context of home loans. Court records show that banks appear to have sought unqualified borrowers, written loans that should go bad, packaged those loans into tranched securities, sold those securities, and retained the CDS contracts (Wyatt, 2011). And since EDCS contracts are derivatives, AIG could not implement a claims process. EDCSs are also vulnerable to the same strategy of "easy money" that could bankrupt the underwriter.

Consequently, to address this problem on the purchaser side of an EDCS, the EDCS **must** be required to transfer with the underlying investment if it is sold off early by a venture bank[6]. It is vital that this principle not be violated or the system will come crashing down sooner or later. It is also important that EDCSs never be sold that can be disconnected from the underlying investment by any other means.

A primary fraud strategy for an EDCS purchaser is to make investment loans and instead of trying to minimize losses on the bottom 80% of their portfolio – the fraudulent venture banker could intentionally wreck lesser performing companies to trigger default faster, and increase the net losses to the underwriter. This results in EDCS contracts requiring more features than the hoary CDS contract. The EDCS must have triggers that the underwriter can use, and it must end with a clawback lien that creates a claim on the assets of the receiver (Hanley 2017; Hanley 2018).

---

6 Venture banks could be required by regulation not to sell either their investments or their EDCS contracts, however, again, I favor flexibility because ability to sell should improve overall market stability as long as this specific perverse incentive problem is addressed.

### 5.9 EDCS triggers

The call side of the EDCS is the underwriter's side. The put side of the EDCS is the venture bank's side. For a trigger, except for the final exit or bankruptcy of the investment, either the qualified holder of the call side of the EDCS can pay the value of the note and take 100% of the equity, or else accept more equity or a new cash payment so that the trigger will not be exercised. Alternatively, after being informed that the trigger event has occurred, the call side of the EDCS could simply allow it to go by.

My preferred conception is that all of the triggers would be optional to exercise, except for the underlying investment's bankruptcy/shutdown and exit. Other triggers within EDCSs would have time limits. Starting points on time limits can be tied to events to make them more flexible.

For all of these triggers except for final exit or bankruptcy, the underwriter can pay off the note (exercise a right to call) and take over the venture bank's insured interest in the company, or else require another payment from the buyer to prevent exercise.

#### *5.9.1 Failure to pay premiums*

Should a EDCS purchaser fail to pay a premium without agreement of the EDCS holder, then the equity in the investment immediately transfers to the EDCS holder, and the buyer's accounts can be raided to collect any penalties in the EDCS. There needs to be a clear mechanism for notifications, what constitutes a valid payment, and reasonable provisions for inability to pay due to factors beyond the control of the purchaser. For instance, it may be in the interest of the underwriter to negotiate some other form of payment, such as an increase in exit equity, if the inability to pay is the result of a temporary banking crisis or other unusual market condition.

#### *5.9.2 Bankruptcy filing by the company invested in*

Bankruptcy of the underlying investment is an automatic trigger that puts the EDCS holder's interests ahead of any other party.

#### *5.9.3 Receipt of an offer to buy, where the venture bank wishes to turn the offer down*

Startup companies receive offers to buy. It is common to turn down the first offer, or even several offers. Mulcahy, et al. identified failure to accept good offers (usually the first offer) as a significant cause of lower returns on investment in venture-capital funds. An EDCS contract might not have this feature, however, that might entail a higher premium or equity. If this feature were present, the underwriter could have the option to take possession of the equity, accept the buyout offer, and pay the venture bank its half of the equity that is above the EDCS value.

#### *5.9.4 Failure to inform in a timely manner that a trigger event has occurred*

Without the investor/investment providing this information, basic trust is violated. It is inevitable that some venture bank will fail to inform by oversight. Either way, this is a trigger that allows taking possession of the venture bank's equity stake in some pre-negotiated manner. My conception is that this would entail taking possession of 50% of the equity immediately, and disposing of it as the underwriter sees fit.

#### *5.9.5 Exit of the investment by the venture bank with a gain*

This always transfers to the holder of the call side of the note the fraction of the equity defined by the EDCS. No payout is made by the underwriter.

### 5.10 The clawback lien – key to removing perverse incentive to defraud: 77/23

When an EDCS contract loss payout is made to a venture bank, the final stage of the derivative instrument is the establishment of a clawback lien against the assets of the venture bank. The clawback lien has first position for payment by the venture bank. The clawback lien only applies to the difference between the insured investment loan amount and the valuation of the equity accepted by the underwriter. It is important that underwriters allocate a fair value to the equity they acquire in the insurance swap because this is the incentive for venture banks to maximize the value of every investment in their portfolio.

There are three ways I see reasonable for an underwriter to operate.

A. *Simple clawback without a claims process*. In this option, the insurer pays the cash value of the EDCS, and takes 100% of the equity in the investment. The clawback specifies that the venture bank must pay back 77% (or some other regulated or negotiated value) of the insurance payment, with interest, when the portfolio of companies this investment loan belongs to exits. Until that time, the bank has full use of the 77% they will pay back later. The venture bank retains the 23% free and clear.

B. *Clawback with a claims process*. Here, the clawback specifies 100% payback to start. The insurer performs a claims process at their option on the investment over a 90 day period, and the lien has a contractual clause that gives the underwriter transparency into the entire history of the investment loan and the investment that the venture bank holds. The underwriter now owns the entire equity of the investment, and it should have special right of audit of the investment, provided that it does not interfere unreasonably with operations. Should the underwriter determine that a violation of its terms has occurred, it may demand the 23% reserve share from the venture bank and recover 100% of its payout at portfolio exit. If there is no violation, then the venture bank releases claim on the 23%, and the 77% clawback is owed.

C. *Clawback with claims for some limited fraction of the portfolio*. An underwriter can monitor its overall operations by using a statistical quality control method to randomly examine a limited number of the payouts it makes. The fraction could be predetermined against the venture bank's portfolio. This method should result in significant cost savings for the underwriters, as claims examinations could be quite involved.

One may ask why a clawback lien should exist at all if most of it will be given back to the underwriter? This is a matter of cash and asset flow.

The purpose of the EDCS is to satisfy banking regulators that the insurance policy will pay off the loan in full should there be a shortfall in the investment return. The payout will remain with the venture bank for the duration of the portfolio it belongs to. So the venture bank will have use of the underwriter's full payment until exits close out of the investment portfolio. There will probably be instances of venture banks that have small portfolios and none of their investments pan out. This would mean that this bank would either fail, and all of its assets would go to the underwriter, leaving the VBU itself to pay off any shortfalls, or it could be required to combine its portfolio with others within the VBU as part of its contract. This is the logic of the VBU which operates one bank but contains many smaller venture bank operations within it that it runs operations for.

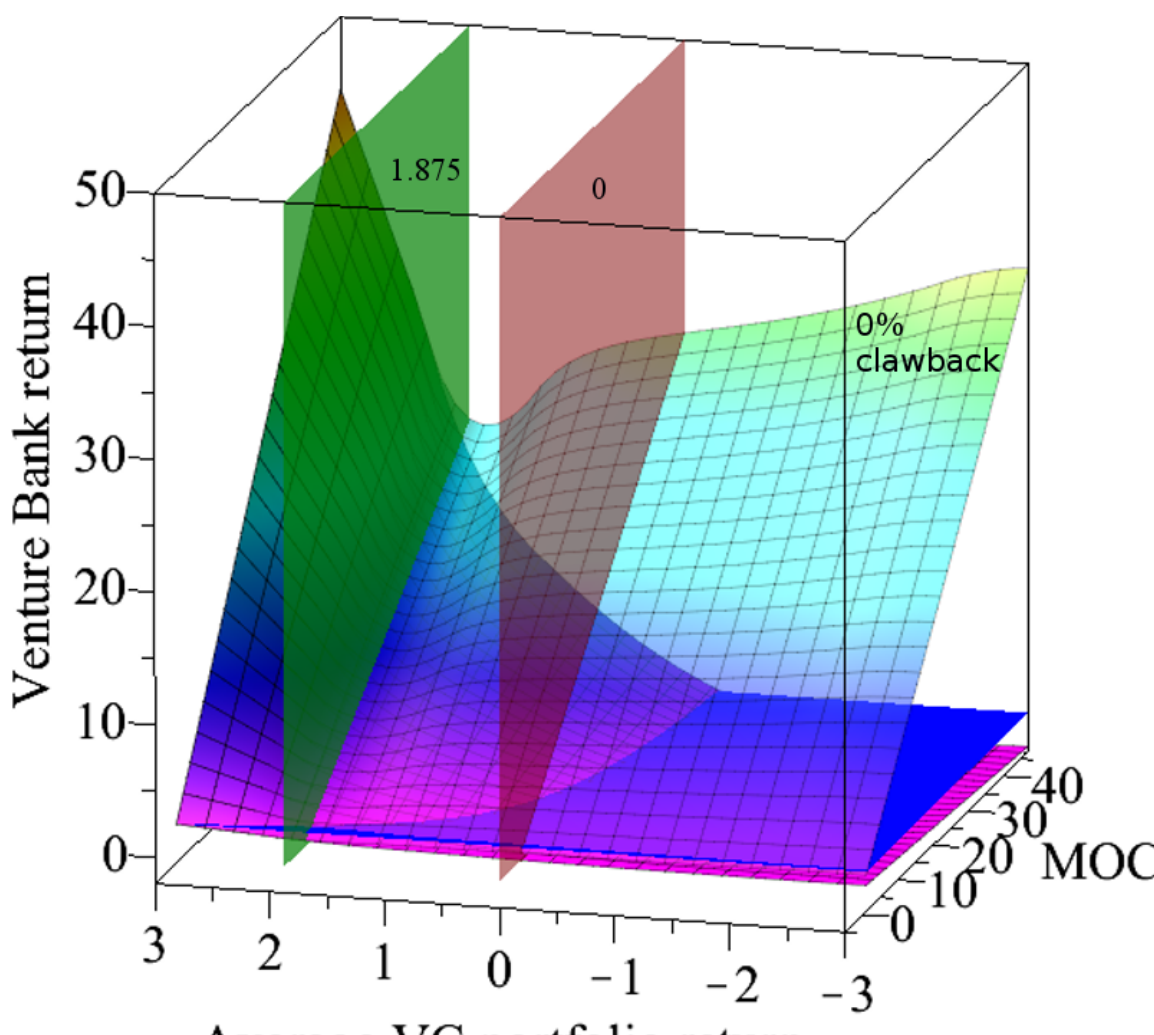

*Figure 18: Graph of venture bank returns with and without clawback. The trough in the 0% clawback graph is where the Kauffman portfolio lands. Reproduced from fig 11A (Hanley, 2018)*

From figure 18 it should be obvious that any fiduciary that sees that their best efforts at a large portfolio will do worse than any other strategy, including crashing everything, will attempt to increase returns by sabotaging investments. A perverse incentive similar to this is what caught AIG in the global financial crisis. The second factor that sealed AIG's fate was that they had not constructed a claims process that could work within a derivative environment.

#### *5.10.1* ***Claim area: Failure of reasonable diligence by venture bank***

To have a case of diligence failure requires formalization of areas of proper diligence and matters that require disclosure. In general, these would include failure to disclose: clear violations of science, physics, causality or mathematics; whether there is clearance to practice (IP clearance) and if not, what potential problems could exist[7]; reasons for believing the team has the skills to carry out the venture; background of relevance – qualifications and experience; criminal charges or convictions of significance; range of market size; market reach strategy feasibility; and future value range projection.

The intention here is not to create a set of "gotcha" clauses. Entrepreneurship is an art, not a science, and there are often gray areas. Sometimes a technology is not fully understood. For instance, how SSRI drugs exert their effect is still a matter of debate, however, clearly these drugs do have value. There will be instances where a technology that is not fully understood does not pan out, but there was enough reason to think it could that it was worth the try. Holding or lacking academic/formal qualifications does not mean that someone can or cannot build a company that delivers a technology.

---

7 Clearance to practice is a gray area with intellectual property in many cases because of overlapping patents, and patents that in some cases should not have been issued. Emphasis should be on clarity of what is known, and straightforward discussion of possible problems and cost to handle them. (e.g. A budget of *$X* per potential litigation.)

Conversely, a claimed technology that depends on using the spin of photons would not be debatable, because we know that photons don't have spin. Similarly, if the total market is relatively small, it would not be debatable that figures that overshoot it by a multiple of 2 or more are unlikely. If some demographic will cost too much to reach by any known method, that would also not be debatable.

The intent is to provide a recourse in the case of egregious failure by the venture bankers to understand the venture that is being invested in, and failure to disclose known risks to the underwriter. This should be different from the case when a venture capitalist makes the occasional mistake, or things don't work out as planned for reasons beyond their control. In addition, it is not uncommon for a venture to find out it can make money in a way it did not plan on, and the original target doesn't pan out. Pivots are a sign of a good team.

Alternatively, an underwriter and venture-bank could agree to not have such a clause to simplify matters.

#### *5.10.2* ***Loss of key personnel***

During important periods, the loss of key personnel can lead to failure. Depending on disclosure, and whether it is possible to get key person insurance for a reasonable cost, this could potentially be grounds for a claim if it was not disclosed. This type of loss may or may not include malfeasance by personnel. (e.g. filing of criminal charges, or similar "morals clause" factors.)

### 5.11 EDCS operations

An EDCS can be written after, or simultaneously with, an investment made by a venture bank in a company. The venture bank's collateral is the equity in the company received for their investment. An underwriter may write an EDCS that encumbers some fraction of venture bank equity or all of it[8].

---

8 There is also the possibility of selling bare call EDCS instrument shares. These could be sold by the venture-bank at a discount. This instrument would get no equity in a trigger scenario. The instrument would collect some equity value at closeout.

In practice, I expect that the EDCS and investment loan would usually be made simultaneously, and that normally 100% of the venture bank equity would be encumbered/covered.

### 5.12 Secondary market

The conceptual structure of EDCS instruments could generate interest on the call side. On the call side (the underwriter), sales of such instruments by the original issuer to qualified buyers would allow those buyers to profit by accepting follow-on payments from purchasers on the put side. They could also seek to profit by collecting on the equity promised by the EDCS buyer at exit.

Defining bite-size shares for the call side, would allow the participation of members of the public in the secondary market who otherwise could not afford the payout risk. In my conception, a controlling interest of 30% or more of the total call valuation should be retained by the underwriter, or else sold to an exercise competent entity. The public buyer would have access to the make-up of the portfolio of companies that the EDCSs insure, or when buying a share of a single EDCS, the company would be defined. This would improve the underwriter's cash flow and lower its liabilities.

In concept, it could be possible to structure things so that the call side investors have voting rights exercised online that could operate if they had a quorum. An alternative simpler way to set up the voting system would be to have buyers set their default choice for the exercise of the EDCS, when they buy. However, if I were operating the call side as the exercise competent entity, I would want to retain maximum flexibility to respond to conditions, without hindrance. So I recommend that public small share secondary market buyers assign any decision to exercise rights to the exercise competent entity to prevent problems.

The underwriter may package EDCSs into a portfolio, and sell shares in that portfolio to the general public. Institutional, or qualified private investors with, may buy the call side of an EDCS contract and transfer the EDCS in toto to themselves.

The danger of a secondary market is that an underwriter could decide to dump its worst investments early, and keep the better ones. Consequently, I suggest that underwriters be required to offer predetermined fractions of their total portfolios to the public, not to exceed 70% of the total. Underwriters can structure this as shares in portfolios, or they may choose to allow up to 70% of the share of any single EDCS to be sold, leaving the buy decision up to the public. Making the decision across the board will ensure that the public gets fair and equal access to the EDCS instrument shares.

### 5.13 Necessity of open market, or trading on an exchange

There is no open market today in CDS derivatives or anything else similar to EDCSs as defined here. The advantages of such a market are large for EDCS contracts.

- With this market mechanism, it would be possible to establish a public valuation of the underlying venture-bank equities.
- Underwriters would be able to sell off some of their risk, lowering their capital requirements.
- Secondary buyers could buy and sell their side of an EDCS to speculate on start-ups success or failure.
- Analysts could provide analysis and opinion. This would need to be protected in some ways, but the extra transparency should be beneficial overall.

## 6 Summary operation of EDCS instruments

The structure of an EDCS typically comprises:

- Premium payments made yearly[9]. Premium payments may be front-loaded or flat fee per year. Premiums could be fixed or floating for a portfolio.
- Pay-out triggers. Triggers for payout of the insured default amount, penalties, haircuts, and seizure of insured equity.
- Exit trigger. A trigger that delivers some fraction of insured equity upon exit.

Parameters that determine cost of EDCSs are:

- Term – 1 year, extendable for 10 years with secondary extensions up to 15 years.
- Exit equity – A variable value paid to the underwriter on exit. The higher the equity, the lower the premium. The lower the premium the greater the net negative cash flow that the underwriter can experience prior to exit for a specific EDCS.
- Bank rate – Bank rate is used to determine the cost of money to carry the pay-off cost of a EDCS portfolio for the underwriter prior to portfolio exit. Bank rate is also used to calculate the final cost of paying back the clawback liens to the underwriters with interest.

9 Alternatively, premiums could be quarterly, however, my modeling used annual payments.

## 7 Audit of EDCS instruments

EDCS instruments have potential risks that are different than most. Consequently, special audit provisions are needed.

1. Validate that contracts to insure equity in an enterprise are the type of shares claimed and have no issues that would render them problematic at a successful exit.
2. Validate that equity to be potentially delivered is not hypothecated or otherwise encumbered.
3. Validate that venture banks are monitoring the insured companies with valid drop-in visits to ensure that the investments are not fictional, that investments have been put into real enterprises intended to produce goods and/or services. This would include, but not be limited to, validating personnel, seeing premises and products in process, and monitoring milestones. Validate that EDCS underwriters have good, timely information on the underlying investment.
4. Validate that the underwriter has the capacity to monitor and administer EDCSs that are triggered, including decisions on whether to pay out or take over equity, handle equity assets in a timely manner, and be able to track what those assets are.
5. Track EDCSs as to original issuer, purchaser, whether an EDCS is a primary or secondary instrument, and that secondary EDCSs are referenced to primary and vice versa.
6. Validate that EDCS portfolio shares being sold are representative of the whole portfolio of the underwriter. Ensure that underwriters are not packaging their worst investments for the public as a way to improve their earnings, while withholding EDCSs that insure their best performing investments.
7. Ensure that the purchaser has not sold or transferred the investment to another entity without first informing the underwriter, and that the put side of the EDCS has not been separated from the loan-investment it is attached to. Doing so should make the EDCS instrument null and void.
8. Validate that a EDCS portfolio discloses the degree of correlation of the underlying assets. EDCS portfolios should be examined for degree of correlation, and should the degree of correlation of the underlying assets be higher than disclosed, strategies should be pursued to mitigate that risk, or the portfolio risk level modified to disclose it to market exchanges and buyers.

9. It is normal for venture-capital firms to collectively over-invest in sectors that are "hot". However, when "me-too" investors over-populate a sector, it is obvious that many will not survive. An audit will examine sectors in which a EDCS portfolio is invested, and evaluate the multiple of the total available market (TAM) and served available market (SAM) that is collectively invested. The multiples of the potential TAM and SAM relative to investments in the sector should be disclosed by the underwriter as public data.

10. Validate that the terms of EDCS coverage expected by interbank lenders and regulators are being complied with. These terms may vary between venture banks, for instance, based on track record or how new the firm is.

11. Pay attention to the rates of capital gains and corporation taxes, and what is or is not excluded from pre or post-tax categorization. These can have significant effects on equity markets, and historically, regulators have responded in ways that exacerbated or created problems (Gardiner, 2006, Chp. 11).

## 8 Concluding remarks

I hope to get people thinking about new ways to approach venture financing. As a side effect, it should result in a significant economic boost as money creation directed into high-value real-economy ventures occurs.

This method should allow expansion of the meaning of venture capital by allowing longer, and more realistic time horizons than the common 5 year target with the J curve fiction. I also think it would be good for a disruptive shake up of how venture-capital operates. The VBU-EDCS system outlined here could provide a structure that would allow various investment groups to form and pool resources. The process to obtain an EDCS should help to improve the odds of success, and compliance with good practices such as the audit described should protect underwriters from the kind of gaming the system that occurred in the the real estate sector. The development of a strong underwriting business is the most important element of this proposal, and it is the keystone on which the ability to operate rests.

There are inherent difficulties with individual venture-capital firms creating banks for themselves to operate. Aside from the extra overhead and impact on mental bandwidth taking away focus from deal flow, a venture-capital firm tends to make large investments as the deals become available, without a great deal of concern about how much that represents in their total portfolio. However, banks need to diversify and minimize the impact of any one investment on the bank. This is why I created the VBU concept.

The VBU can bridge the operations gap between the more narrowly focused venture-capital firms and the need for the venture banks to have a broad portfolio in order to be healthy. There are regulatory limits in banking on the amount that any one investment represents, typically 10%. By operating multiple firm's funds for them, in quasi-siloed operations within the VBU, the VBU and its underwriting partner can operate over a broad portfolio that represents a large ecosystem of varied investments.

There may be venture capital firms that are larger, with multiple decades of operations experience and a diverse portfolio of investments, that could be candidates for converting their investment operations over to this method, operating their own bank and underwriting.

One could ask, why create venture banking when it is already the case that banks can create new deposits through loan activity? After all, current banks could do so without the burden of private market underwriting, and this insurance mechanism described takes half the equity. The answer to this starts with the observation that banks do not do this today because of regulation. In the regulatory scheme, EDCS contracts (properly implemented) prevent bad behavior by venture banks, while allowing regulators a very light hand.

Much regulation has gone into creating bounds within which banks can operate because of the inherent moral hazard of banking. Requiring underwriters to insure venture loan investments means that an adversarial party will have to agree. An area where venture banking could prove problematic is for finance of finance, because such deals can present excellent returns on paper in short time scales. So leveraged buyouts, securitizations, financing of resale of existing assets, and similar should be proscribed to venture banking to prevent fueling bubbles of the type that created the 2008 crisis (Hanley, 2012). Venture loan underwriting should be directed to investments in enterprises that create new utility value or else improve efficiency.

## 9 Acknowledgements

I want to thank Geoffrey Gardiner for his observation that the EDCS represents a new way to match the term of a loan with the term of funds for the bank. I also thank Steve Keen for critique and advice.

## 10 Glossary

AIG – American International Group. A global insurance company providing insurance products to commercial, institutional and individual customers. They also provide mortgage insurance and credit default swap (CDS) contracts.

BACPA – Bankruptcy Abuse Prevention and Consumer Protection Act of 2005. For these purposes, BACPA strengthened the rights of derivative holders to collect immediately.

Basel accords – There are three sets of banking regulations set by the Basel Committee on Bank Supervision. These are known as Basel I, Basel II and Basel III.

Call – In options trading, a call contract gives the holder the right to buy an asset at a pre-negotiated price for some time period. For a EDCS, it signals the side of the EDCS that collects the assets of an investment in return for payment of the insurance value to the purchaser, and collects equity at exit of the investment.

CDS – Credit Default Swap. The purchaser makes premium payments to the underwriter and the contract insures a loan on some asset, typically a real estate loan. If the borrower defaults on the loan, then the purchaser is paid the face value of the contract, and transfers the asset to the underwriter.

EDS – Equity Default Swap. A proposed derivative that covers loans made by venture capitalists as investments, defined in this paper. It functions similarly to a CDS.

EDCS – Equity Default Clawback Swap. This improvement on the EDS that covers loans removes the perverse incentive to create losses. (The arsonist motive.) This EDCS contract is they key to enabling this new type of banking to function.

FRED – Federal Reserve Economic Data.

Haircut – A reduction in the stated value of an asset.

IPO – Initial Public Offering.

IP – Intellectual property. Patents, trademarks and copyrights.

M&A – Merger and Acquisition.

MOC – Multiple of Original Capital. Some amount of money is put into the bank that is its capital. This amount is enlarged by the Basel accords rules into the complete Tier 1 and Tier 2 capital that is used by the bank as reserves. The total outstanding investments divided by the original capital placed in bank Tier 1 reserves is the MOC. See figure 6.

Put – In options trading, a put is a contract that buys the right to force a buyer for your asset to pay a pre-negotiated price. For a EDCS, it is the right to collect the payoff amount of the insurance and turn over the equity of an insured investment.

TAM – Total Available Market.

VBU – Venture Bank Utility. This is a proposed new entity that handles the banking operations for a set of venture banks. See figure 5.